\documentclass[aps,preprint,prd,showpacs,nofootinbib]{revtex4}
\usepackage{amsmath,amssymb,amsfonts,array}
\usepackage{bm}
\usepackage{cases}
\usepackage{dcolumn}
\usepackage{graphicx}
\usepackage{hyperref}
\usepackage{subfigure}
\usepackage{wasysym}
\usepackage{xcolor}

\definecolor{dullred}{rgb}{0.706,0.208,0.192}
\definecolor{darkred}{rgb}{0.545,0,0}
\definecolor{MaroonC}{rgb}{0,0.502,0.502}
\definecolor{dullblue}{rgb}{0,0.298,0.49}
\definecolor{blue3}{RGB}{31, 119, 180}
\definecolor{dullpurple}{rgb}{0.431,0.188,0.534}
\definecolor{darkgreen}{rgb}{0.075,0.302,0.047}
\definecolor{darkergreen}{rgb}{0,0.196,0.125}
\definecolor{darkergreen2}{rgb}{0,0.294,0.188}

\hypersetup{colorlinks=true,
	breaklinks=true,
	pdfstartview=Fit,
	linkcolor=blue,
	citecolor=blue,
	urlcolor=blue}

\def\be{\begin{equation}}
\def\ee{\end{equation}}
\def\ba{\begin{eqnarray}}
\def\ea{\end{eqnarray}}

\def\nn{\nonumber}

\begin{document}
	
%\begin{flushleft}
%	{\footnotesize
%		UCAS/2019/01
%	}
%\end{flushleft}

\title{Null energy condition violation during inflation and pulsar timing array observations}

\author{Gen Ye$^{1,2}$}
\email[]{ye@lorentz.leidenuniv.nl}
\author{Mian Zhu$^{3}$}
\email[Corresponding author:~]{mian.zhu@uj.edu.pl}
\author{Yong Cai$^{1}$}
\email[Corresponding author:~]{caiyong@zzu.edu.cn}
\affiliation{$^1$ School of Physics and Microelectronics, Zhengzhou University, Zhengzhou, Henan 450001, China}
\affiliation{$^2$ Institute Lorentz, Leiden University, P.O. Box 9506, Leiden 2300 RA, The Netherlands}
\affiliation{$^3$ Faculty of Physics, Astronomy and Applied Computer Science, Jagiellonian University, 30-348 Krakow, Poland}

\begin{abstract}

Recently, evidence of stochastic gravitational wave background (SGWB) signals observed by pulsar timing array (PTA) collaborations, has prompted investigations into their origins. We explore the compatibility of a proposed inflationary scenario, incorporating an intermediate null energy condition (NEC)-violating phase, with the PTA observations. The NEC violation potentially amplifies the primordial tensor power spectrum, offering a promising explanation for PTA observations.
Numerical analyses, primarily focused on NANOGrav's 15-year results, reveal the model's compatibility with PTA data. Notably, the model predicts a nearly scale-invariant GW spectrum in the mHz frequency range, which sets our scenario apart from other interpretations predicting a red primordial GW spectrum on smaller scales.

\end{abstract}
%\pacs{98.80.-k, 98.80.Cq, 04.50.Kd}

\maketitle
\tableofcontents

\section{Introduction}
\label{sec:intro}

Gravitational Waves (GWs) are promising tools for exploring the physics of the early universe. Observations of GWs from binary pulsar systems \cite{Detweiler:1979wn} and the merger of binary black holes \cite{LIGOScientific:2016aoc} have ushered in the era of GW detection.
Recently, pulsar timing array (PTA) collaborations, including NANOGrav \cite{NANOGrav:2023hvm,NANOGrav:2023gor}, EPTA \cite{EPTA:2023fyk}, PPTA \cite{Reardon:2023gzh}, and CPTA \cite{Xu:2023wog}, announced compelling evidence of a signal consistent with stochastic gravitational wave
background (SGWB) at the reference frequency $f = 1~ \textnormal{yr}^{-1}$.
Shortly after the release of the PTA results, various studies emerged regarding the possible origin of the observed signal, see, e.g., \cite{Addazi:2023jvg,Kitajima:2023vre,Athron:2023mer,Ghoshal:2023fhh,Yang:2023aak,
Bai:2023cqj,Ellis:2023dgf,Huang:2023chx,Franciolini:2023pbf,Li:2023yaj,Kitajima:2023cek,Datta:2023vbs,
Fujikura:2023lkn,Vagnozzi:2023lwo,Ellis:2023tsl,Wang:2023len,Guo:2023hyp,Han:2023olf,
Shen:2023pan,Zu:2023olm,Bi:2023tib,Wang:2023ost,Broadhurst:2023tus,Cai:2023dls,
Depta:2023qst,Bian:2023dnv,Franciolini:2023wjm,Du:2023qvj,Wu:2023hsa,Yi:2023mbm,
Liang:2023fdf,Anchordoqui:2023tln,Battista:2023znv,DeFalco:2023djo,Li:2023bxy,Xiao:2023dbb,Zhang:2023lzt,Liu:2023ymk,
Inomata:2023zup,Ghosh:2023aum,Niu:2023bsr,Konoplya:2023fmh,DiBari:2023upq,
Wang:2023sij,Ye:2023xyr,Balaji:2023ehk,Zhang:2023nrs,Jiang:2023gfe,Zhu:2023lbf,
An:2023jxf,Hooper:2023nnl,King:2023ayw,Maji:2023fhv,Datta:2023xpr,
HosseiniMansoori:2023mqh,Frosina:2023nxu,He:2023ado,Ellis:2023oxs,Kawasaki:2023rfx,
Kawai:2023nqs,Bhattacharya:2023ysp,Huang:2023zvs,Lozanov:2023rcd,Bernardo:2023zna,Chen:2023bms,Ahmadvand:2023lpp,Chowdhury:2023xvy,Aghaie:2023lan,Choudhury:2024one,Choudhury:2023fjs,Choudhury:2023fwk, Choudhury:2023hfm,Choudhury:2013woa}.

Primordial GWs \cite{Grishchuk:1974ny,Starobinsky:1979ty,Rubakov:1982df}, which are the tensor fluctuations generated quantum mechanically in the very early universe, serve as an important possible source of the SGWB. Detecting the primordial GW background would provide valuable physical insights into the origin and evolution of the universe. It is intriguing to explore whether recent observations by PTA can be interpreted as signals of the primordial GWs with a blue-tilted power spectrum. Recently, a lot of research has been conducted in this direction, see, e.g., \cite{Vagnozzi:2023lwo,Jiang:2023gfe,Zhu:2023lbf,Borah:2023sbc,Choudhury:2023kam,
Ben-Dayan:2023lwd,Oikonomou:2023qfz,Datta:2023xpr}.

Inflation \cite{Guth:1980zm,Linde:1981mu,Albrecht:1982wi,Starobinsky:1980te}, as a leading paradigm for the very early universe, elegantly addresses the horizon and flatness problems of the Big Bang cosmology. It also predicts a nearly scale-invariant power spectrum of scalar perturbations, which is consistent with the cosmic microwave background (CMB) observations. Furthermore, the conventional slow-roll inflationary scenario also predicts a nearly scale-invariant power spectrum of primordial GWs. However, the PTA observations suggest a highly blue tensor spectrum with a spectral index $n_T = 1.8 \pm 0.3$ \cite{NANOGrav:2023hvm,Vagnozzi:2023lwo}, which implies the necessity of new physics beyond the scope of conventional slow-roll inflation at certain scales (see e.g. \cite{Piao:2004tq,Baldi:2005gk,Piao:2006jz,Kobayashi:2010cm,Kobayashi:2011nu,Endlich:2012pz,
Cai:2014uka,Gong:2014qga,Cannone:2014uqa,Wang:2014kqa,Kuroyanagi:2014nba,Cai:2015yza,Cai:2016ldn,Wang:2016tbj,Fujita:2018ehq,
Kuroyanagi:2020sfw,Akama:2020jko,Akama:2023jsb,Giare:2020plo,Giare:2022wxq} for explorations in generating a blue-tilted tensor spectrum within the inflationary scenario).

The violation of the null energy condition (NEC) may play a crucial role in the very early universe (see e.g. \cite{Rubakov:2014jja} for a review). It has been demonstrated that a fully stable NEC violation can be realized in the ``beyond Horndeski'' theories \cite{Cai:2016thi,Creminelli:2016zwa,Cai:2017tku,Cai:2017dyi,Kolevatov:2017voe,
Ilyas:2020zcb,Ilyas:2020qja,Zhu:2021ggm}, see also \cite{Libanov:2016kfc,Kobayashi:2016xpl,Ijjas:2016vtq,Dobre:2017pnt,Zhu:2021whu,Cai:2022ori}.
Basically, the violation of NEC implies an increase in the Hubble parameter $H$ (i.e., ${d H/dt}>0$). Given that the power spectra of scalar and tensor perturbations are proportional to $H^2$, an intermediate NEC violation during inflation leads to intriguing phenomena of observational interest at certain scales, including: 1) a significant enhancement in the power spectrum of primordial GWs \cite{Cai:2020qpu,Cai:2022nqv}, 2) a notable amplification of the parity-violation effect in primordial GWs \cite{Cai:2022lec} (see also \cite{Zhu:2023lhv}), 3) the generation of primordial black holes with masses and abundances of observational interest \cite{Cai:2023uhc}, along with the associated scalar-induced GWs. Remarkably, the scenario proposed by \cite{Cai:2020qpu} naturally yields a broken power-law power spectrum, which may potentially be consistent with observations from both CMB and PTA, see also \cite{Jiang:2023gfe} for recent studies.

In this paper, we examine the predicted primordial GW background from the model proposed by \cite{Cai:2020qpu} in light of observations from PTA, with a specific focus on the NANOGrav signal.
Our paper is organized as follows. In Sec. \ref{sec:model}, we provide a brief overview of the background dynamics in our model. Sec. \ref{sec:PGW01} introduces the power spectrum and its parameterization as predicted by our model. In Sec. \ref{sec:num01}, we perform numerical comparisons between the primordial GW background predicted by our model and the NANOGrav signal. Sec. \ref{sec:conclusion231007} is dedicated to our conclusion.

\section{Intermediate NEC violation during inflation}
\label{sec:model}

\subsection{A brief review of the scenario}

\begin{figure}[htp]
    \centering    \includegraphics[width=0.55\linewidth]{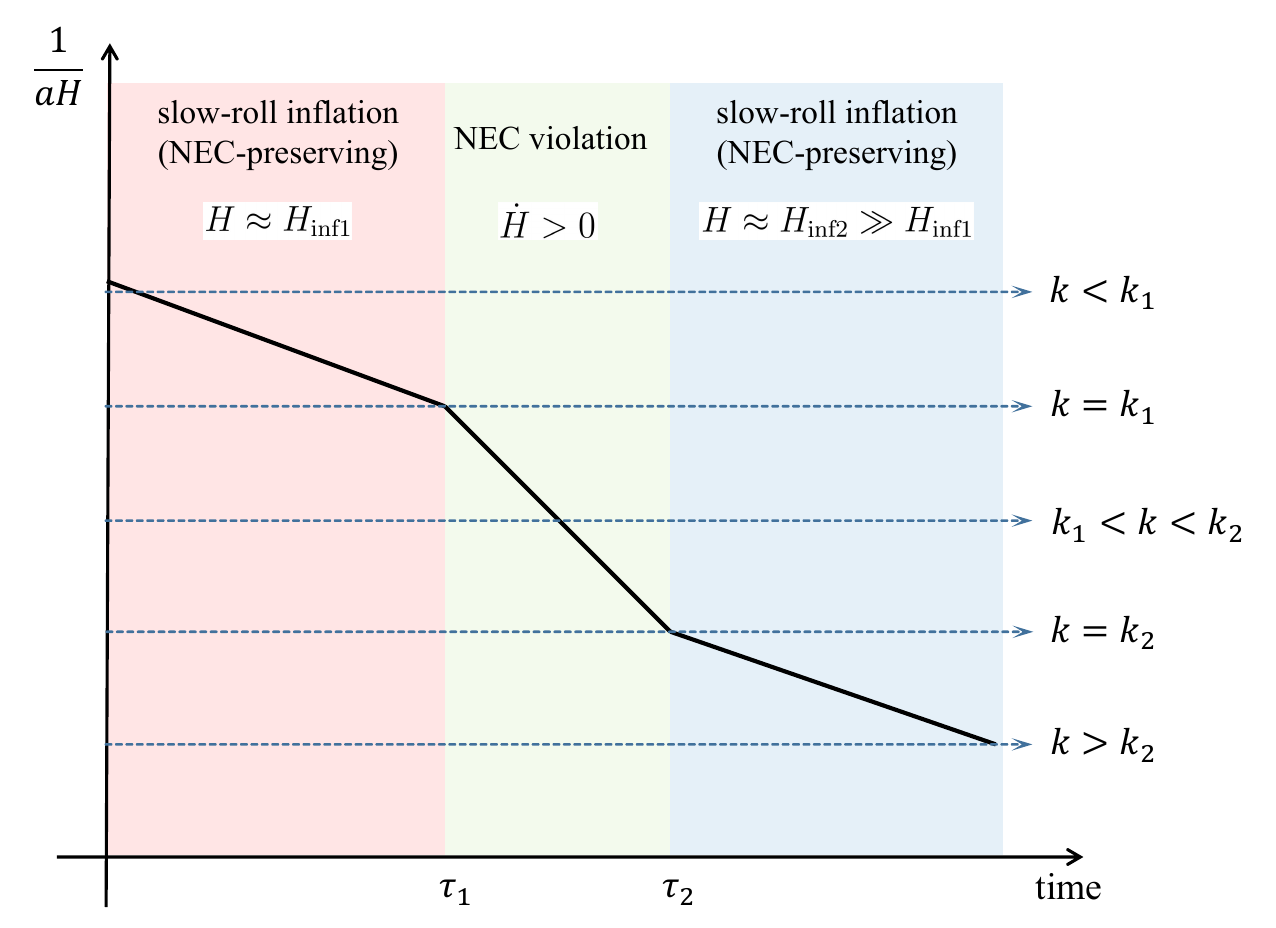}
    \caption{An illustration of the evolution of the conformal Hubble horizon (represented by the solid black lines) and the perturbation modes (depicted as dashed blue lines with arrows) in our scenario.}
    \label{fig:BGandPert-1}
\end{figure}

In our scenario \cite{Cai:2020qpu}, the universe begins with an initial phase of slow-roll inflation characterized by a Hubble parameter $H=H_\text{inf1}$. Subsequently, it transitions into a second phase of slow-roll inflation with a significantly larger Hubble parameter, denoted as $H=H_\text{inf2}$, after passing through an intermediate stage of NEC violation.
During the evolution, the comoving Hubble horizon (i.e., $a^{-1}H^{-1}$) decreases over time, signifying the exit of perturbation modes from the horizon as the universe undergoes accelerated expansion, see Fig. \ref{fig:BGandPert-1} for an illustration.\footnote{The horizon of scalar perturbations, as defined by $\sqrt{z_s/z_s''}$, can differ from the Hubble horizon, especially during the transition between different phases. Here, we have simplified the analysis by not considering the specific details of this transition.} After crossing the horizon, these perturbation modes remain in the super-horizon regime until they re-enter the horizon during the subsequent radiation-dominated or matter-dominated expansion era.

For the scalar (or tensor) perturbation modes that exit the horizon during the first stage of slow-roll inflation (i.e., $k < k_1$), their power spectrum is nearly scale-invariant, making it consistent with the observations of temperature anisotropy in the CMB. Similarly, the power spectrum of perturbation modes that exit the horizon during the second stage of slow-roll inflation (i.e., $k > k_2$) is also nearly scale-invariant, but it has a significantly larger amplitude. The scale-invariance of the tensor power spectrum at large scales ensures the absence of a highly suppressed tensor-to-scalar ratio $r$, and consequently a highly suppressed slow-roll parameter in canonical single-field slow-roll inflation. At small scales, the scale-invariance of the scalar and tensor power spectra prevents them from growing to $\mathcal{O}(1)$, thus preserving the validity of perturbation theory at higher frequencies.

The power spectrum of perturbation modes that exit the horizon during the NEC-violating phase (i.e., $k_1 < k < k_2$) is blue-tilted ($n_s>0$ or $n_T>0$). Namely, the violation of the NEC enhances both the scalar and tensor power spectrum by increasing the Hubble parameter $H$, see \cite{Cai:2020qpu,Cai:2022nqv,Cai:2022lec,Cai:2023uhc}.
Significantly, at the intermediate scales of the scalar power spectrum, the blue tilt and oscillatory features, particularly around the scale corresponding to the beginning of the second inflationary phase, lead to intriguing phenomena of observational interest, including the generation of primordial black holes and the associated scalar-induced GWs \cite{Cai:2023uhc}. Simultaneously, the power spectrum of tensor perturbations (or the primordial GWs), also exhibits a blue tilt and possesses oscillatory features around the transition scales \cite{Cai:2020qpu,Cai:2022nqv,Cai:2022lec}.
These distinctive features set our scenario apart from other single-field primordial black hole formation scenarios.
The combination of primordial black holes, scalar-induced GW signals, and primordial GWs provides a valuable avenue for studying the violation of the NEC during inflation, especially in the era of multi-messenger and multi-band observations.

Our scenario can be realized with the EFT action (see e.g., \cite{Cai:2016thi,Cai:2017tku,Cai:2022nqv})
\ba
S&=&\int
d^4x\sqrt{-g}\Big[ {M_{\rm P}^2\over2} R-\Lambda(t)-c(t)g^{00}
\nn\\
&\,&\qquad\qquad\quad +{M_2^4(t)\over2}(\delta g^{00})^2-{m_3^3(t)\over2}\delta
K\delta g^{00}  + {\tilde{m}_4^2(t)\over
	2}R^{(3)}\delta g^{00}\Big]\,,
\label{action02}
\ea
where $\delta g^{00}=g^{00}+1$, $R^{(3)}$ is the Ricci scalar on the 3-dimensional spacelike hypersurface, $\delta K=K-3H$, $K$ is the extrinsic curvature.
The time-dependent functions $c(t)$ and $\Lambda(t)$ determine the background evolution, with the relations $c(t) = -M_{\rm P}^{2} \dot{H}$ and $\Lambda(t) = M_{\rm P}^{2} (\dot{H} + 3 H^{2})$. The functions $M_2^4(t)$, $m_3^3(t)$ and $\tilde{m}_4^2(t)$ can be determined or constrained based on the condition that the scalar perturbations are in agreement with observations.

Notably, the operator $R^{(3)}\delta g^{00}$ plays a crucial role in preventing scalar perturbations from becoming unstable when the NEC is violated, as demonstrated in \cite{Cai:2016thi,Creminelli:2016zwa,Cai:2017tku}.
The covariant form of action (\ref{action02}), as discussed in \cite{Cai:2017dyi,Kolevatov:2017voe}, falls under the category of ``beyond Horndeski'' theory. Nonetheless, the propagation of primordial GWs is exactly the same as that in general relativity at quadratic order.
In this paper, we will not delve into the specific formulations of these coefficient functions or the intricacies of the model construction. Instead, we will employ a simplified parameterization of the background evolution for our scenario. Consequently, we can establish a parameterization for the power spectrum of primordial GWs.

\subsection{Parametrization of background}

We begin with a flat Friedmann-Lemaitre-Robertson-Walker universe described by the metric
\begin{equation}
    ds^2 = -dt^2 + a(t)^2 d\vec{x}^2 = a(\tau)^2 (-d\tau^2 +  d\vec{x}^2) ~,
\end{equation}
where $t$ is the cosmic time and $\tau$ is the conformal time, related by $dt = ad\tau$. Throughout this paper, we will use a dot to denote $d/dt$ and a prime to denote $d/d\tau$. We will also define $H\equiv \dot{a}/a$, ${\cal H}\equiv a'/a$.

An inflationary stage is commonly characterized by quasi-de Sitter expansion, where the scale factor approximately evolves as $a \propto |\tau|^{-1}$ for $\tau<0$. Additionally, we will parameterize our NEC-violating stage with a power-law scale factor, i.e., $a(\tau) \propto |\tau|^n$.\footnote{An exponential parameterization of the scale factor during the NEC-violating stage is employed in Appendix \ref{App:exp01}.}
The full-scale factor can be then parametrized as a piecewise function of $\tau$:
\begin{equation}
\label{eq:atau}
    a_j(\tau) = a_j (\tau_j) \left( \frac{\tau - \tau_{R,j}}{\tau_j - \tau_{R,j}} \right)^{\frac{1}{\epsilon_j - 1}} ~,\quad \tau<\tau_j \,,
\end{equation}
%\begin{equation}
%\label{eq:atau}
%    a_j(\tau) = a_j (\tau_j) \left( \frac{\tau - \tau_{R,j}}{\tau_j - \tau_{R,j}} \right)^{\frac{1}{\epsilon_j - 1}} ~,~ \mathcal{H}_j (\tau) \equiv \frac{a_j^{\prime}}{a_j} = \frac{1}{\epsilon_j - 1} \frac{1}{\tau - \tau_{R,j}} ~.
%\end{equation}
where $\tau_j$ is the conformal time at the end of phase $j$, $j=1,2,3$ corresponds to the first inflation stage, the NEC-violating stage and the second inflation stage, respectively; ${\tau}_{R,j} =\tau_{j}- (\epsilon_{j}-1)^{-1} \mathcal{H}^{-1}(\tau_j)$ is the integration constant and $\epsilon_j=-\dot{H}/H^2$ is treated as constant during phase $j$. 
Since phases 1 and 3 are assumed as slow-roll inflation, we will set $\epsilon_1 \approx \epsilon_3\approx 0$ for simplicity. As for the NEC-violating phase (i.e., phase 2), we have $\epsilon_2 < 0$, which indicates $-1<n=(\epsilon_2-1)^{-1}<0$. A specific design of such a model can be found in \cite{Cai:2020qpu}.

The provided parametrization overlooks the details of transitions between different phases, including the specific variations of $\epsilon$ near the beginning or the end of the NEC-violating phase. The dynamics of the transition might depend on the specific model. For simplicity, we will describe the transition physics using the matching condition, ensuring the continuity of the scale factor $a$ and its first derivative $a'$ at $\tau_1$ and $\tau_2$. For our purpose, such a simplification will not make a qualitative difference.

Using Eq. (\ref{eq:atau}) and denoting the beginning and the end time of the NEC-violating stage to be $\tau_1$ and $\tau_2$, respectively, the continuity of $a$ gives
\begin{equation}
    a_1(\tau_1) = a_2(\tau_2) \left( \frac{\tau_1 - \tau_{R,2}}{\tau_2 - \tau_{R,2}} \right)^{\frac{1}{\epsilon_2 - 1}} ~,~ a_2(\tau_2) = a_3(\tau_3) \left( \frac{\tau_2 - \tau_{R,3}}{\tau_3 - \tau_{R,3}} \right)^{-1} ~.
\end{equation}
The continuity of $a'$ or $\mathcal{H}$ enables us to define the following quantities
\begin{equation}
    \bar{\mathcal{H}}_1 \equiv \mathcal{H}(\tau_1) ~,~ \bar{\mathcal{H}}_2 \equiv \mathcal{H}(\tau_2) ~.
\end{equation}
With the help of \eqref{eq:atau}, we can solve the integration constants as
\begin{equation}
\label{eq:Hjunction}
    \tau_{R,1} = \tau_1 + \bar{\mathcal{H}}_1^{-1} ~,~ \tau_{R,2} = \tau_1 - \frac{\bar{\mathcal{H}}_1^{-1}}{\epsilon_2 - 1} = \tau_2 - \frac{\bar{\mathcal{H}}_2^{-1}}{\epsilon_2 - 1} ~,~ \tau_{R,3} = \tau_2 + \bar{\mathcal{H}}_2^{-1} ~.
\end{equation}
Obviously, the consistency of \eqref{eq:Hjunction} requires
\begin{equation}
    (\tau_2 - \tau_1 ) (1-\epsilon_2) = \bar{\mathcal{H}}_1^{-1} -  \bar{\mathcal{H}}_2^{-1} > 0 ~.
\end{equation}
In terms of the scale factor and Hubble parameter;
\begin{equation}
   \frac{a_2(\tau_2)}{a_2(\tau_1)} = \left( \frac{\bar{\mathcal{H}}_2}{\bar{\mathcal{H}}_1} \right)^{\frac{1}{1-\epsilon_2}} ~,~ \frac{H_2}{H_1} = \left( \frac{\bar{\mathcal{H}}_2}{\bar{\mathcal{H}}_1} \right)^{\frac{\epsilon_2}{\epsilon_2 - 1}} ~,
\end{equation}
where we have defined $H_2 \equiv \bar{\mathcal{H}_2}/a_2(\tau_2)$ and $H_1 \equiv \bar{\mathcal{H}_1}/a_2(\tau_1)$ to be the Hubble parameter at $\tau = \tau_2$ and $\tau = \tau_1$.

\section{Primordial gravitational waves}\label{sec:PGW01}

\subsection{Tensor perturbations and mode functions}

Since the gravity sector is minimally coupled to the matter sector, the quadratic action for tensor perturbation is simply
\begin{equation}
    S_T^{(2)} = \int d\tau d^3x \frac{a^2}{8} \left[ \gamma_{ij}^{\prime 2} - (\partial_k \gamma_{ij})^2 \right] ~,
\end{equation}
where we have set the propagation speed of tensor perturbation to be unity (i.e., the speed of light), and $M_\mathrm{P}=1$. In the momentum space, the dynamical equation for tensor perturbation is
\begin{equation}
\label{eq:MSeq}
    u_k^{\prime \prime} + \left( k^2 - \frac{a^{\prime \prime}}{a} \right) u_k = 0 ~,
\end{equation}
where $u_k \equiv \gamma_k^{\lambda} a/2$ is the mode function and $\lambda = +,\times$ represent two different polarizations.

The parameterization of scale factor, i.e., Eq. \eqref{eq:atau}, gives
\begin{equation}
    \frac{a^{\prime \prime}_j}{a_j} = \frac{\nu_j^2 - 1/4}{(\tau - \tau_{R,j})^2} ~,\quad ~ \nu_j \equiv \frac{1}{2} + \frac{1}{1 - \epsilon_j} ~.
\end{equation}
Note that $\epsilon_2<0$ would result in $1/2 < \nu_2 < 3/2$.
As a result, in each phase, the general solution to Eq. \eqref{eq:MSeq} can be expressed in terms of the Hankel function as
\begin{equation}
    u_{k,j}(\tau) = \frac{\sqrt{\pi (\tau_{R,j} - \tau)}}{2} \left\{ \alpha_j H_{\nu_j}^{(1)} [k(\tau_{R,j} - \tau)] + \beta_j H_{\nu_j}^{(2)} [k(\tau_{R,j} - \tau)] \right\} ~,
\end{equation}
where $H_{\nu_j}^{(1)}$ and $H_{\nu_j}^{(2)}$ are the first and
second kind Hankel functions of the $\nu_j$-th order, respectively; $\alpha_j$ and $\beta_j$ are $k$-dependent coefficients.

For simplicity, we have assumed $\epsilon_1 \approx \epsilon_3 \approx 0$, which indicates  $\nu_1 \approx \nu_3 \approx 3/2$ and the Hankel functions are simply
\begin{equation}
    H_{3/2}^{(1)}(x) = \sqrt{\frac{2}{\pi}} \frac{e^{ix}}{x^{3/2}} (-i-x) ~,~
    H_{3/2}^{(2)}(x) = \left[ H_{3/2}^{(1)}(x) \right]^{\ast} ~,
\end{equation}
where the asterisk denotes complex conjugation.
We impose the Bunch-Davis vacuum initial condition, which gives $|\alpha_1| = 1$ and $\beta_1 = 0$. The other coefficients $\alpha_j$ and $\beta_j$ for $j=2$, $3$ can be determined with the matching method, which requires the continuities of $u_k$ and $u_k^\prime$ at the transition surface $\tau = \tau_1$ and $\tau = \tau_2$.

More explicitly, we are interested in the final tensor spectrum, which is relevant to $|\alpha_3 - \beta_3|$, namely
\begin{equation}
    P_T \equiv \frac{4k^3}{\pi^2} \frac{|u_{k,3}^2|}{a^2} = \frac{2H(\tau_3)^2}{\pi^2} |\alpha_3-\beta_3|^2 ~.
\end{equation}
We present
\begin{align}
\label{eq:alphabeta3}
 & \quad \frac{8 ( \alpha_3-\beta_3 )}{\pi(1-2\nu_2)\sqrt{x_1 y_1}} \nonumber =
H^{(1)}_{\nu_2}(y_2)\left[\left(i-\frac{1}{x_1} \right)H^{(2)}_{\nu_2-1}(x_2) + H^{(2)}_{\nu_2}(x_2)  \right]\sin y_1 \\
& \nonumber + H^{(1)}_{\nu_2-1}(y_2) \left[\left(i-\frac{1}{x_1} \right)H^{(2)}_{\nu_2-1}(x_2) + H^{(2)}_{\nu_2}(x_2) \right] \left(\cos y_1 - \frac{\sin y_1}{y_1} \right) - \\
& \left[\frac{ix_1-1}{x_1}H^{(1)}_{\nu_2-1}(x_2) + H^{(1)}_{\nu_2}(x_2) \right] \left[ H^{(2)}_{\nu_2}(y_2)\sin y_1 + H^{(2)}_{\nu_2-1}(y_2) \left(\cos y_1 - \frac{\sin y_1}{ y_1} \right) \right] ~,
\end{align}
see \cite{Cai:2022nqv}, where we employ
\begin{equation}
    x_1 \equiv \frac{k}{\bar{\mathcal{H}}_1} ~,~ x_2 = \left( \nu_2 - \frac{1}{2} \right) x_1 ~;~ y_1 \equiv \frac{k}{\bar{\mathcal{H}}_2} ~,~ y_2 = \left( \nu_2 - \frac{1}{2} \right) y_1 ~,
\end{equation}
and an overall phase factor has been neglected.

\subsection{Parametrization of the tensor spectrum}

The tensor spectrum provided by \eqref{eq:alphabeta3} is complicated, necessitating further simplification. To interpret the PTA data through primordial GWs from our model, the amplitude of primordial GWs must be enhanced from the order of $\mathcal{O}(10^{-11})$, the upper bound on the amplitude of $P_T$ constrained by CMB observations, to $10^{-3}$ within the PTA range. Given that in our scenario, $n_T < 2$, the NEC violating stage must persist for scales spanning over four orders of magnitude, i.e., $k_2 > 10^4 k_1$. Consequently, we can safely make the approximation $k_2 \ll k_1$.

Next, we observe that the Hankel function behaves as a pure phase factor in the sub-horizon region and as a power-law function in the super-horizon region. This asymptotic behavior allows us to derive approximate expressions for modes exiting the horizon in different stages. For example, for modes exiting the horizon during the second inflation stage, where $k \gg k_2$, leading to $y_1, y_2, x_1, x_2 \gg 1$, each Hankel function can be approximated as a pure phase function, resulting in
\begin{equation}
P_T \simeq P_{T,2} \equiv \frac{2H_2^2}{\pi^2} ~, \quad k > k_2 ~.
\end{equation}
Similarly, modes exiting the horizon during the first inflation stage satisfy $k < k_1$, for which we have
\begin{equation}
P_T \simeq P_{T,1} \equiv \frac{2H_1^2}{\pi^2} ~, \quad k < k_1 ~.
\end{equation}

For perturbation modes that exit the horizon during the NEC-violating stage, the approximate tensor power spectrum is given by
\begin{equation}
P_T \simeq \frac{\Gamma^2(\nu_2)}{\pi} \left( \frac{2\nu_2 - 1}{4} \right)^{1-2\nu_2} \left( \frac{k}{k_2} \right)^{3-2\nu_2} ~, \quad k_1 < k < k_2 ~,
\end{equation}
see \cite{Cai:2022nqv} for more details.
This spectrum shows a blue tilt ($n_T=3-2\nu_2>0$) within the range $k_1 < k < k_2$ in our scenario. Since $1/2 < \nu_2 < 3/2$, we have $0<n_T<2$. 
Remarkably, when employing an exponential parameterization for $a_2(\tau)$, we reach $n_T=2$, as explicitly demonstrated in Appendix \ref{App:exp01}.

The above treatment encounters challenges near the transition scales $k \simeq k_1$ and $k \simeq k_2$, where a simple expansion of the Hankel function through its asymptotic behavior is not applicable. However, in the vicinity of the scale $k \simeq k_1$, the corresponding tensor spectrum is expected to be too small for detection in the near future, allowing us to safely overlook the transition feature in this region. Conversely, near the scale $k \simeq k_2$, the tensor spectrum is sufficiently large for potential detection.
For $k \gg k_1$, all Hankel functions with arguments $x_1$, $x_2 \gg 1$ exhibit the asymptotic behavior $H_{\nu}^{(1)}(x) \propto \exp [i(x- \nu \pi/2 - \pi/4)]$. This characteristic enables us to capture the features of $P_T$ around $k \simeq k_2$ while simplifying the formulation of $P_T$.

In light of this, we can parametrize $P_T$ as
\begin{equation}
P_{T} = P_{T,1} + \frac{\pi}{4}(2-n_T) \frac{k}{k_2}|g(k)|^2 P_{T,2} ~,\label{PT1211}
\end{equation}
where the auxiliary function is defined as
\begin{equation} \label{eq:g}
g = H_{\frac{3-n_T}{2}}^{(1)} \left[ \frac{2-n_T}{2} \frac{k}{k_2} \right] \sin \frac{k}{k_2} + H_{\frac{1-n_T}{2}}^{(1)} \left[ \frac{2-n_T}{2} \frac{k}{k_2} \right]\left( \cos \frac{k}{k_2} - \frac{k_2}{k} \sin \frac{k}{k_2} \right) ~.
\end{equation}
It is important to note that we have replaced $\nu_2$ by the corresponding tensor spectral index $n_T=3-2\nu_2$ in the NEC violating stage. In the limit of $k\ll k_1$, $P_T\approx P_{T,1}$ since the second term in Eq. (\ref{PT1211}) becomes negligible compared to $P_{T,1}$. Conversely, for $k\gg k_1$, the second term in Eq. (\ref{PT1211}) becomes dominant. Consequently, it ensures that the features around $k\simeq k_2$ are well captured.
While this formulation may sacrifice accuracy around the first transition scale $k\simeq k_1$, such a compromise has minimal impact on our interests.

The primordial tensor power spectrum $P_T$ can be converted to the observed  GW energy spectrum by \cite{Turner:1993vb}
\begin{equation}
\begin{aligned}
\Omega_{\rm{GW}} &= \frac{k^2}{12a_0^2 H_0^2}\left[\frac{3\Omega_mj_l(k\tau_0)}{k\tau_0}\sqrt{1.0 + 1.36\frac{k}{k_{eq}}+2.50\left(\frac{k}{k_{eq}}\right)^2}\right]^2P_T\\
&\simeq \frac{1}{24}\left(\frac{k}{H_0}\right)^2\left[\frac{3\Omega_m}{(k\tau_0)^2}\sqrt{1.0 + 1.36\frac{k}{k_{eq}}+2.50\left(\frac{k}{k_{eq}}\right)^2}\right]^2P_T
\end{aligned}
\end{equation}
where $H_0=67.8\, {\rm km/s/Mpc}$, $\tau_{0}=1.41\times10^{4}$ Mpc, $a_0=1$, $k_{\text{eq}}=0.073\,\Omega_{\text{m}} h^{2}$
Mpc$^{-1}$, $\Omega_{\rm m}$ is the density fraction of matter today, and the wavenumber relates to the frequency as $k=2\pi f$.

\section{Numerical results}\label{sec:num01}

In this section, we confront the parameterized spectrum \eqref{PT1211} to the recent PTA data. As explained in the previous section, we are interested in the case $k_1\ll k\lesssim k_2$ where the tensor spectrum is blue. To this end, the first term in \eqref{PT1211} is much smaller than the second one, thus we can safely neglect $P_{T,1}$. With this simplification, the theoretical spectrum \eqref{PT1211} is fully parameterized by three parameters $\{P_{T,2}, \ n_T, \ f_c\}$, where we have defined the transition frequency from the NEC-violating stage to the second inflationary stage as $f_c\equiv 2\pi k_2$. We performed Monte Carlo Markov Chain (MCMC) analysis varying $\{P_{T,2}, \ n_T, \ f_c\}$ plus the pulsar and nuisance parameters against the most recent public NANOGrav 15yr dataset \cite{NANOGrav:2023gor}. Following NANOGrav, we assume two models on the spatial correlation of the signal, uncorrelated common-spectrum red noise (CURN) and Hellings-Down (HD).\footnote{CURN assumes spatially uncorrelated signal while HD assumes that the signal at different pulsars has a spatial correlation described by the Hellings-Down curve \cite{Hellings:1983fr}. The later is the expected spatial correlation of SGWB. See \cite{NANOGrav:2023gor} for details.}. Theoretically, our model predicts $n_T<2$. The validity of the perturbation theory requires $P_{T,2}<1$. See Table.\ref{tab:prior} for our prior choice of the model parameters.

\begin{figure}[htbp]
	\centering 	\includegraphics[scale=2,width=0.75\textwidth]{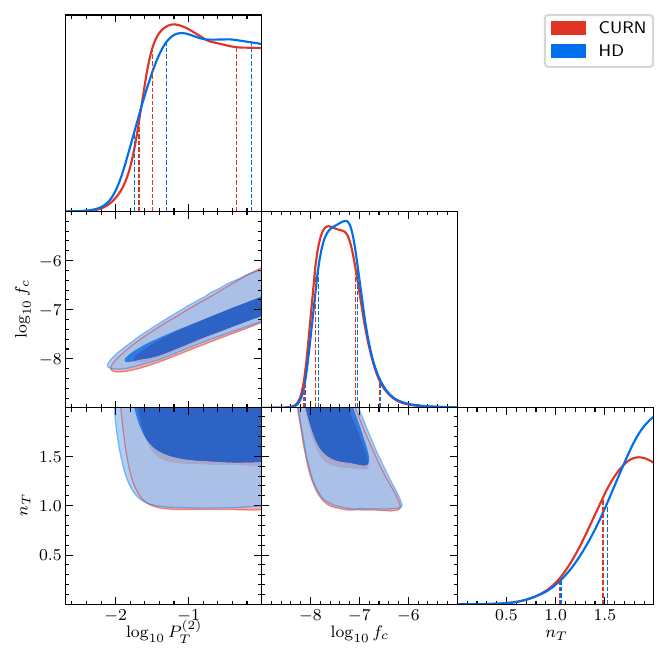}
	\caption{68\% and 95\% posterior distributions of model parameters.}
	\label{fig230802-1}
\end{figure}

\begin{figure}[htbp]
    \subfigure[~~$\Omega_\text{GW}h^2$]{\includegraphics[width=.48\textwidth]{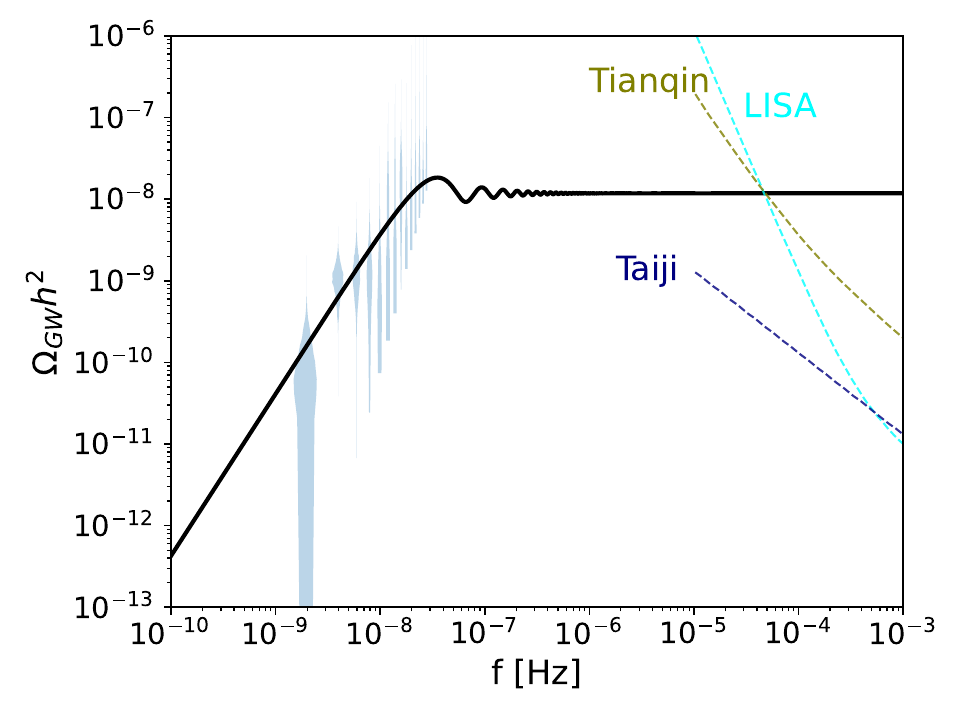} }
    \subfigure[~~Excess timing]{\includegraphics[width=.48\textwidth]{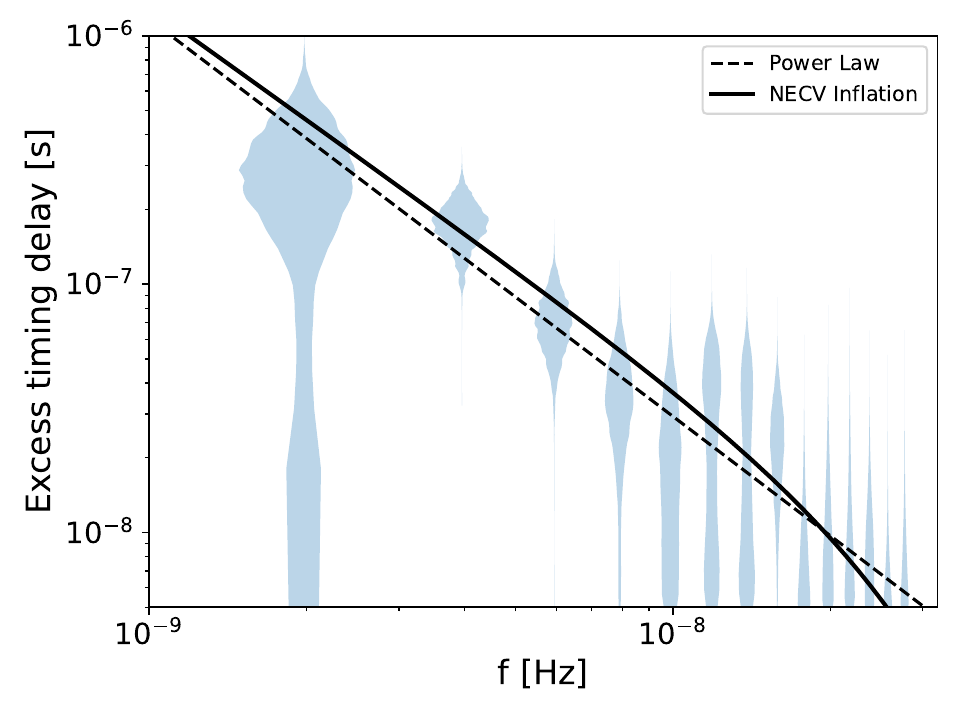} }
    \caption{Theoretical spectra with violin data points from NANOGrav 15yr.} \label{fig230802-2}
\end{figure}

Fig.\ref{fig230802-1} shows the posterior distributions of the tensor spectrum parameters. Data puts lower bounds on $P_{T,2}>10^{-2}$ (95\% C.L.) and $n_T>1.0$ (95\% C.L.). This is to be expected because it is known that the NANOGrav 15yr data includes a $\sim3\sigma$ detection of GW background near $10^{-8}$ Hz with $n_T\sim 1.8$. The distribution of $f_c$ is a peak $-8.1<\log_{10}(f_c/\mathrm{Hz})<-6.6$, but this should not be confused with a detection. Near $f_c$, the spectrum transits from $n_T\sim 2$ to $n_T=0$ with oscillations. Having $f_c>10^{-8}$ Hz means that data does not favor the shape of transition so it is moved to the right of the data constraining region, see also Fig.\ref{fig230802-2}. The upper bound on $f_c$ is due to the prior upper bound on $P_{T,2}$. Because data fixes the amplitude of the spectrum near $10^{-8}$ Hz, $f_c$ and $P_{T,2}$ are strongly positively correlated, as can be seen from the $\log_{10}f_c$ - $\log_{10} P_{T,2}$ contour of Fig.\ref{fig230802-1}. Therefore, a prior upper bound $P_{T,2}<1$ gets translated to an upper bound on $f_c$.

Fig.\ref{fig230802-2} further compares the best-fit theoretical spectrum with violin points measured by NANOGrav. The left panel of Fig.\ref{fig230802-2} illustrates the physical energy spectrum of GW denoted as $\Omega_{GW}h^2$. While the NEC-violating phase satisfactorily accounts for the PTA signal, the tensor spectrum originating from the second inflationary phase also falls comfortably within the detection sensitivity of forthcoming space-based GW observatories like LISA \cite{LISA:2017pwj}, Taiji \cite{Hu:2017mde}, and Tianqin \cite{TianQin:2015yph}. The right panel of Fig.\ref{fig230802-2} zooms into the PTA data-constrained frequency range and plots the PTA timing excess, see \cite{NANOGrav:2023gor} for a detailed definition. Only the first few data points detect GW background signal and the theoretical spectrum (solid black line) fits them as well as a power law (dashed black line). Close to $f_c$, the solid line drops to a smaller $n_T$, which is not favored by data. Therefore, in the MCMC, $f_c$ automatically shifts to a higher frequency than the range constrained by the data, leading to the observed lower bound on $f_c$ as depicted in Fig.\ref{fig230802-1}.

The numeric results confirmed that the NEC-violating model can well explain the GW background signal observed by PTA experiments, but the current data is not precise enough to distinguish it from other models, e.g., a simple power law. Our model also predicts a flat tensor spectrum in the mHz frequency range and is detectable to next-generation space GW detectors.

\begin{table}[htbp]
    \centering
    \begin{tabular}{|c|c|}
    \hline
        Parameter&Prior\\
        \hline
        $\log_{10}P_{T,2}$ & [-5,0] \\
        $n_T$ & $[0,2]$\\
        $\log_{10}f_c$ & [-9,-5]\\
        \hline
    \end{tabular}
    \caption{Prior choice for spectrum parameters in the MCMC analysis.}
    \label{tab:prior}
\end{table}

\section{Conclusions}\label{sec:conclusion231007}

The recent SGWB signals reported by PTA collaborations have unveiled a new frontier in the exploration of gravitational wave physics. Notably, if these PTA signals originate from the primordial universe, it necessitates the presence of new physics beyond the standard slow-roll inflation scenario. This necessity arises because PTA observations indicate a highly blue tensor spectrum with a spectral index of $n_T = 1.8 \pm 0.3$, whereas the conventional slow-roll inflation scenario predicts a nearly scale-invariant tensor spectrum.

An intermediate stage of NEC violation during inflation has the potential to amplify the primordial tensor power spectrum on certain scales, offering a potential explanation for the PTA observations.
In this paper, we explore the primordial GW spectrum within an inflationary scenario featuring an intermediate NEC-violating phase. 
We present parameterizations for the background evolution and the GW power spectrum of our model. Our evaluation of the model's compatibility with PTA data reveals its capability to account for the SGWB signal observed by PTA experiments. Given the consistency of signals observed by various PTA experiments, we focus primarily on the NANOGrav 15-year results in the numerical analysis.

Additionally, our model predicts a nearly scale-invariant GW spectrum in the mHz frequency range, potentially detectable by upcoming space-based GW detectors like LISA, Taiji, and Tianqin. This distinctive characteristic distinguishes our scenario from other interpretations predicting a red primordial GW spectrum on smaller scales.

\section{Discussions}

Intriguingly, Ref. \cite{Cai:2022lec} found that, for the scenario discussed in this paper, the violation of the NEC naturally amplifies the parity-violating effect as well as its observability in primordial GWs, provided the scalar field determining the background evolution is coupled to
a parity-violating term. The wavenumber $k$ corresponding to the maximum of the parity-violating effect is approximately the same as the wavenumber corresponding to the maximum of the power spectrum. This intriguing feature also sets our model apart observationally from other models.

The scale corresponding to the significantly amplified parity-violating effects depends on the scale at which the NEC violation takes place. In this paper, the maximum of the primordial GW power spectrum appears in the vicinity of the PTA band, as shown by the black curve in Fig. \ref{fig230802-2}. Therefore, the parity violation is still notably amplified in the PTA band. 
However, considering the power spectrum presented in Fig. \ref{fig230802-2}, the parity violation is suppressed by the slow-roll condition in the LISA band, as the GW modes in the LISA band are generated during the second slow-roll inflation stage in our scenario. To achieve a notable amplification of the parity-violating effect in the LISA band, it would be necessary for the NEC violation to occur later than what is assumed in the present paper.

In Ref. \cite{Cai:2023uhc}, it has been demonstrated that the NEC violation can naturally enhance the primordial scalar power spectrum at
certain scales, leading to the production of PBHs and scalar-induced GWs of observational interest. The primordial GWs (i.e., tensor perturbations) primarily depend on the Hubble parameter $H$. The scalar-induced GWs are induced by scalar perturbations, depending not only on $H$ but also on $\epsilon\equiv-\dot{H}/H^2$ (or its generalized formulation). Therefore, the relative contribution to the GW background of each depends on the specific model's characteristics, particularly the relative magnitudes of the power spectra of scalar and tensor perturbations.

In this paper, we primarily focus on primordial GWs and have not explicitly calculated the primordial scalar perturbations. We assume that the contribution to the GW background in the PTA band mainly comes from primordial GWs rather than scalar-induced GWs. In fact, for the same background evolution, scalar perturbations exhibit stronger model dependence compared to tensor perturbations. In principle, for models constructed with specific covariant action, we can quantitatively compare the contributions of both to the GW background. This remains a subject for future work.

\acknowledgments

We are grateful to Yi-Fu Cai, Chunshan Lin, Yun-Song Piao, Fei Wang and Yi Wang for stimulating discussions.  Y. C. is supported in part by the National Natural Science Foundation of China (Grant No. 11905224), the China Postdoctoral Science Foundation (Grant No. 2021M692942) and Zhengzhou University (Grant Nos. 32213984, 32340442). G.Y. is supported by NWO and the Dutch Ministry of Education, Culture and Science (OCW) (grant VI.Vidi.192.069).  M.Z. is supported by grant No. UMO 2021/42/E/ST9/00260 from
the National Science Centre, Poland. PTA data analysis is performed using the \texttt{PTArcade} software \cite{andrea_mitridate_2023_8106173, Lamb:2023jls, ellis_2020_4059815, enterprise-ext}. We acknowledge the use of the computing
server {\it Arena317}@ZZU.

\appendix
\section{Power spectrum from an exponential parameterization of the scale factor during the NEC-violating phase}\label{App:exp01}

In this section, we explicitly calculate the power spectrum $P_T$ using an exponential parameterization for $a_2(\tau)$, representing the scale factor during the NEC-violating stage. This parametrization is frequently used in the study of bouncing cosmology, which also has an NEC-violating phase, see e.g., \cite{Cai:2012va}.

The first and second inflationary stages, denoted by $i=1$ and $3$ respectively, are still parameterized by Eq. (\ref{eq:atau}), with the assumption that $\epsilon_1\approx \epsilon_3 \approx 0$. Therefore, we have
\begin{equation}
    a_1(\tau) = a_1(\tau_1) \frac{\tau_1 - \tau_{R,1}}{\tau - \tau_{R,1}} ~,~ \mathcal{H}_1(\tau) = - \frac{1}{\tau - \tau_{R,1}} ~, ~ \tau\leq \tau_1
\end{equation}
and
\begin{equation}
    a_3(\tau) = a_3(\tau_3) \frac{\tau_3 - \tau_{R,3}}{\tau - \tau_{R,3}} ~,~ \mathcal{H}_3(\tau) = - \frac{1}{\tau - \tau_{R,3}} ~,~ \tau_2\leq \tau \leq \tau_3
\end{equation}
If the NEC-violating phase is very brief, we can approximate the Hubble parameter as a linear function. As a result, $a_2(\tau)$ can be parameterized by an exponential formulation as 
\begin{equation}
    a_2(\tau) = a_1 (\tau_1) e^{\mathcal{H}_1(\tau_1) (\tau-\tau_1) + \frac{1}{2} \gamma^2 (\tau - \tau_1)^2} ~,~ \mathcal{H}_2(\tau) = \mathcal{H}_1(\tau_1) + \gamma^2 (\tau - \tau_1) ~,
\end{equation}
where $\tau_1\leq\tau\leq\tau_2$.

The parameterization of the background can be related to the characteristic scales: 
\begin{equation}
    k_1 = \mathcal{H}_1(\tau_1) = (\tau_{R,1} - \tau_1)^{-1} ~,~ k_2 = \mathcal{H}_3(\tau_2)  = (\tau_{R,3} - \tau_2)^{-1} ~,
\end{equation}
where $k_1$ is the mode that crosses the horizon at the end of the first inflationary phase, $\tau = \tau_1$, and $k_2$ is the mode that crosses the horizon at the beginning of the second inflationary phase, $\tau = \tau_2$.

The continuity of $a$ and $\mathcal{H}$ at $\tau_1$ is already incorporated into the parametrization. Therefore, we focus on ensuring continuity at $\tau_2$, which requires that $a_3(\tau_2)=a_2(\tau_2)$ and $\mathcal{H}_3(\tau_2)=\mathcal{H}_2(\tau_2)$, i.e.,
\begin{equation}
    \frac{a_3(\tau_2)}{a_1(\tau_1)} = e^{\mathcal{H}_1(\tau_1) (\tau_2 - \tau_1) + \frac{1}{2} \gamma^2 (\tau_2 - \tau_1)^2} ~,~ - \frac{1}{\tau_2 - \tau_{R,3}} = \mathcal{H}_1(\tau_1) + \gamma^2 (\tau_2 - \tau_1) ~.
\end{equation}
In terms of $k_1$ and $k_2$, and considering the condition $k_1 \ll k_2$, we can express
\begin{equation}
k_2 = k_1 + \gamma^2 (\tau_2 - \tau_1) \simeq \gamma^2 (\tau_2 - \tau_1) ~,
\end{equation}
leading to
\begin{equation}
\label{eq:a3a1}
\tau_2 - \tau_1 \simeq \frac{k_2}{\gamma^2} , \frac{a_3(\tau_2)}{a_1(\tau_1)} \simeq e^{\frac{k_2}{2\gamma^2}(k_2 + 2k_1)} \simeq e^{\frac{k_2^2}{2\gamma^2}} ~.
\end{equation}

For the first inflationary phase (i.e., $\tau<\tau_1$), the mode function can be given as
\begin{equation}
    u_{k,1}(\tau) = \frac{\sqrt{\pi (\tau_{R,1}-\tau)}}{2} H_{3/2}^{(1)} [k(\tau_{R,1} - \tau)] = \frac{e^{ik (\tau_{R,1} - \tau)}}{\sqrt{2} k^{3/2} (\tau_{R,1} - \tau)} [-i - k(\tau_{R,1} - \tau)] ~,
\end{equation}
where we have taken the vacuum initial condition. It can be further simplified to
\begin{equation}
    u_{k,1}(\tau) = - \frac{k+ik_1}{\sqrt{2}k^{3/2}} e^{\frac{ik}{k_1}} ~,~~ \tau<\tau_1 ~.
\end{equation}

In the NEC-violating phase, we have $\epsilon \equiv 1 - {\mathcal{H}^\prime}/{\mathcal{H}^2} \ll -1$, indicating $\mathcal{H}^{\prime} \gg \mathcal{H}^2$ for nearly the entire phase. As a result, ${a^{\prime \prime}}/{a} = \mathcal{H}^2 + \mathcal{H}^{\prime} \simeq \mathcal{H}^{\prime} = \gamma^2 $. Consequently, the dynamical equation simplifies to
\begin{equation}
u_{k,2}^{\prime \prime}(\tau) - \omega_k^2 u_{k,2}(\tau) = 0 , \omega_k \equiv \sqrt{\gamma^2 - k^2} ~.
\end{equation}
The parameter $\omega_k$ turns imaginary for $k > \gamma$ and real for $k < \gamma$. Yet, we will primarily focus on analyzing the dynamics of perturbations for $k < \gamma$ as it should be adequate for our purposes. In this scenario, $\omega_k$ is real, and the general solution takes the form
\begin{equation}
u_{k,2} = c_{1} e^{\omega_k \tau} + c_{2} e^{-\omega_k \tau} ~.
\end{equation}

The mode function for the second inflationary phase can be expressed as
\begin{equation}
u_{k,3}(\tau) = c_3 \sqrt{\frac{\pi (\tau_{R,3} - \tau)}{2}} J_{\frac{3}{2}} [k(\tau_{R,3} - \tau)] + c_4 \sqrt{\frac{\pi (\tau_{R,3} - \tau)}{2}} Y_{\frac{3}{2}} [k(\tau_{R,3} - \tau)] ~,
\end{equation}
where $J$ and $Y$ are Bessel functions. In a more explicit form, this can be written as
\begin{equation}
u_{k,3}(\tau) = c_3 \left( \frac{\sin z}{z} - \cos z \right) - c_4 \left( \frac{\cos z}{z} + \sin z \right) ~,
\end{equation}
where $z\equiv k(\tau_{R,3} - \tau)$.
On super-horizon scales, the dominating term is $c_4$, making it sufficient to evaluate $c_4$. The tensor spectrum on these scales, expressed in terms of $c_4$, is given by
\begin{equation}
P_T = \frac{4k^3}{\pi^2} \frac{|u_{k,3}^2|}{a^2} = \frac{2H_{\text{inf,2}}^2}{\pi^2} 2k|c_4|^2 ~.
\end{equation}

We match the perturbation at $\tau_1$ and $\tau_2$ for fluctuations with $k<\gamma$, resulting in
\begin{align}
\label{eq:c4}
     c_4 & \nonumber = \frac{k^2 + (ik - k_1)(k_1+\omega_k)}{2\sqrt{2}k^{7/2} \omega_k} e^{\frac{\omega_k}{k_2}} \Bigg[ k^2 \sin \frac{k}{k_2} \\
     & + (k_2 - \omega_k)\left( k \cos \frac{k}{k_2} - k_2 \sin \frac{k}{k_2} \right) \Bigg] + (\omega_k \to -\omega_k) ~.
\end{align}
A quick check: for modes crossing the horizon in the first inflation phase, we have $k<k_1 \ll k_2$. Moreover, the largeness of $|\epsilon|$ implies $|\epsilon|(\tau_1) = 1 - \frac{\gamma^2}{k_1^2} \gg 1$, or $\gamma \gg k_1 \geq k$, and thus $\omega_k \simeq \gamma$.
A series expansion around $k/k_2 \ll 1$ gives
\begin{equation}
    c_4 \simeq \frac{(ik-k_1)}{6\sqrt{2}k_2^2k^{\frac{1}{2}}} e^{\frac{\gamma}{k_2}} \left( 2k_2 + \gamma \right) + (\gamma \to -\gamma) ~.
\end{equation}
Therefore, for $k \ll k_1$, the power spectrum is
\begin{align}
\label{eq:PTinf1}
    P_{T,\text{inf1}} & \nonumber \simeq \frac{2H_{\text{inf,2}}^2}{\pi^2} \times  \frac{k_1^2}{k_2^2} \left[ \frac{1}{6} e^{\frac{\gamma}{k_2}} \left( 2 + \frac{\gamma}{k_2} \right) + (\gamma \to -\gamma) \right]^2 \\
    & \nonumber = \frac{2H_{\text{inf,1}}^2}{\pi^2} \frac{a_1^2}{a_2^2} \left[ \frac{1}{6} e^{\frac{\gamma}{k_2}} \left( 2 + \frac{\gamma}{k_2} \right) + (\gamma \to -\gamma) \right]^2 \\
    & \simeq \frac{2H_{\text{inf,1}}^2}{\pi^2} \left[ \frac{1}{6} \left( 2 + \frac{\gamma}{k_2} \right) + \frac{a_1}{a_2} \left( 2 - \frac{\gamma}{k_2} \right) \right]^2 \simeq P_{T,\text{inf1}} \left( \frac{1}{3} + \frac{\gamma}{6k_2} \right)^2 ~,
\end{align}
which is scale-invariant, as expected. The equation \eqref{eq:PTinf1} indicates that the super-horizon modes generally experience a modification factor during the NEC-violating phase, aligning with results from bouncing cosmology.

Modes with $k > k_2$ and $k > \gamma$ remain sub-horizon at $\tau = \tau_2$ and cross the horizon in the second inflationary phase. Therefore, we can directly write down their corresponding tensor spectrum:
\begin{equation}
\label{eq:PT2}
    P_{T,\text{inf2}} \simeq \frac{2H_{\text{inf,2}}^2}{\pi^2} ~.
\end{equation}
To explain the PTA result, we aim for $P_{T,\text{inf2}} \simeq \mathcal{O}(10^{-4})$. The CMB constraint gives $P_{T,\text{inf1}} \simeq \mathcal{O}(10^{-12})$. Consequently, we can estimate $H_{\text{inf,2}}/H_{\text{inf,1}} > \mathcal{O}(10^4)$. This estimation provides us with an estimate for $\gamma$, i.e.,
\begin{equation}
    \frac{H_{\text{inf,2}}}{H_{\text{inf,1}}} = \frac{\mathcal{H}_2}{\mathcal{H}_1} \frac{a_1}{a_2} > \mathcal{O}(10^4) ~\to~ \frac{k_2}{k_1} e^{\frac{k_2^2}{2\gamma^2}} > \mathcal{O}(10^4) ~.
\end{equation}
The exponential function is sensitive to the value of $k_2/ \gamma$, which suggests that $k_2$ and $\gamma$ have the same order of magnitude. For example, if $k_2/\gamma = 10$, it would imply an amplification of the Hubble parameter by $e^{50}$, an implausible result. Therefore, we could reasonably approximate $P_{T} = P_{T,\text{inf2}}$ for $k > \gamma$, accepting a minor loss of precision within a narrow scale range.

\begin{figure}[htp]
    \centering
    \includegraphics[width=0.55\textwidth]{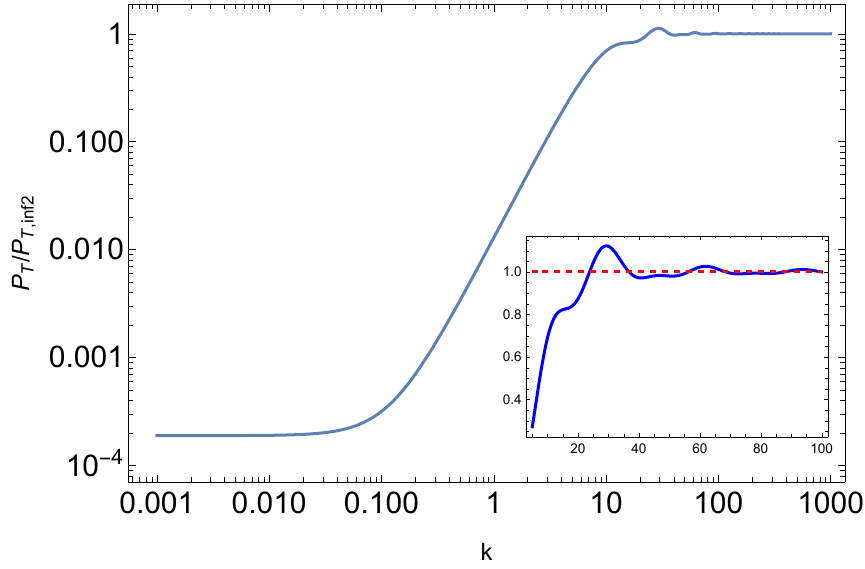}
    \caption{The tensor power spectrum with $k_1 = 0.12$, $k_2 = 10$, $\gamma = 8$.}
    \label{fig:PTexp}
\end{figure}

Obtaining the tensor spectrum for $k > \gamma$ from \eqref{eq:c4} via analytical continuation on $\omega_k$ might raise a potential issue. The expression for $c_4$ displays an apparent discontinuity at $k = \gamma$ and$\omega_k = 0$. 
Looking into this issue in the limit of $k \to \gamma$, we find
\begin{equation}
    \lim_{k \to \gamma} c_4 \simeq \frac{\left( \gamma^2 - k_2^2 \right) \sin \frac{\gamma}{k_2} + \gamma k_2 \cos \frac{\gamma}{k_2}}{\sqrt{2} \gamma^{\frac{3}{2}}k_2} ~.
\end{equation}
Therefore, $c_4$ is well-defined at $k = \gamma$. 
Moreover, for $k \gg \gamma$,
\begin{equation}
    c_4 \simeq \frac{1}{2\sqrt{2}k^{\frac{3}{2}} } \left\{ \frac{e^{\frac{i|\omega_k|}{k_2}}}{i |\omega_k|} \left[ k^2 \sin \frac{k}{k_2}- i|\omega_k| \left( k \cos \frac{k}{k_2} - k_2 \sin \frac{k}{k_2} \right) \right] + c.c. \right\} ~.
\end{equation}
Using $|\omega_k| \simeq k$, it can be further simplified to 
\begin{equation}
    |c_4| = \frac{1}{\sqrt{2}k^{\frac{1}{2}}} \left| 1 + \mathcal{O} \left( \frac{k_2}{k} \right)\right| \to P_{T} \simeq  \frac{2H_{\text{inf,2}}^2}{\pi^2} = P_{T,\text{inf2}} ~.
\end{equation}
Therefore, we can evaluate $P_T$ with $k > \gamma$ through the analytical continuation of $\omega_k$, which we verified numerically and illustrate in Fig. \ref{fig:PTexp}. 

In conclusion, we could adopt the following parametrization for $P_T$:
\begin{equation}
    P_T = \frac{H_{\text{inf,2}}^2}{2\pi^2} |f(k)^2| ~,
\end{equation}
where
\begin{align}\label{eq:fk01}
     f(k) & \nonumber = \frac{k^2 + (ik - k_1)(k_1+\omega_k)}{k^{3} \omega_k} e^{\frac{\omega_k}{k_2}} \Bigg[ (k_2 - \omega_k)\left( k \cos \frac{k}{k_2} - k_2 \sin \frac{k}{k_2} \right)  \\
     & + k^2 \sin \frac{k}{k_2} \Bigg] + (\omega_k \to -\omega_k) ~,~ \omega_k \equiv \sqrt{\gamma^2 - k^2} ~.
\end{align}

How does this relate to PTA observations? As we can see from Fig. \ref{fig:PTexp}, the tensor spectrum is blue-tilted in the region $k_1 \ll k \ll k_2$. Since $\gamma$ is comparable to $k_2$, we also have $\gamma \gg k$ such that $\omega_k \simeq \gamma$. The expression (\ref{eq:fk01}) simplifies to
\begin{equation}
    |f(k)| \simeq \frac{k}{3 k_2^2} \left[ (2 k_2+\gamma) e^{\frac{\gamma}{k_2}} + (2 k_2-\gamma) e^{-\frac{\gamma}{k_2}}\right] ~,
\end{equation}
which predicts a blue spectrum with $n_T = 2$ in the range $k_1 \ll k \ll k_2$. This result is in agreement with that in the main text in the limit $n \to \infty$, as the power-law parametrization $a_2 \propto |\tau|^n$ approaches an exponential function.

However, when $k \sim \gamma$ and $\gamma \neq k_2$, we also observe a blue spectrum from Fig. \ref{fig:PTexp}. Yet, estimating the spectral index analytically becomes challenging in this scenario as we lack a reliable method to estimate $\omega_k$.

%%%%%%%%%%%%%%%%%%%%%%%%%%%%%%%%%%
% bibliography

%\bibliographystyle{plain}
\bibliography{nec}

\providecommand{\href}[2]{#2}\begingroup\raggedright\begin{thebibliography}{100}

\bibitem{Detweiler:1979wn}
S.~L. Detweiler, ``{Pulsar timing measurements and the search for gravitational
  waves},'' \href{http://dx.doi.org/10.1086/157593}{{\em Astrophys. J.}
  {\bfseries 234} (1979) 1100--1104}.

\bibitem{LIGOScientific:2016aoc}
{\bfseries LIGO Scientific, Virgo} Collaboration, B.~P. Abbott {\em et~al.},
  ``{Observation of Gravitational Waves from a Binary Black Hole Merger},''
  \href{http://dx.doi.org/10.1103/PhysRevLett.116.061102}{{\em Phys. Rev.
  Lett.} {\bfseries 116} no.~6, (2016) 061102},
  \href{http://arxiv.org/abs/1602.03837}{{\ttfamily arXiv:1602.03837 [gr-qc]}}.

\bibitem{NANOGrav:2023hvm}
{\bfseries NANOGrav} Collaboration, A.~Afzal {\em et~al.}, ``{The NANOGrav 15
  yr Data Set: Search for Signals from New Physics},''
  \href{http://dx.doi.org/10.3847/2041-8213/acdc91}{{\em Astrophys. J. Lett.}
  {\bfseries 951} no.~1, (2023) L11},
  \href{http://arxiv.org/abs/2306.16219}{{\ttfamily arXiv:2306.16219
  [astro-ph.HE]}}.

\bibitem{NANOGrav:2023gor}
{\bfseries NANOGrav} Collaboration, G.~Agazie {\em et~al.}, ``{The NANOGrav 15
  yr Data Set: Evidence for a Gravitational-wave Background},''
  \href{http://dx.doi.org/10.3847/2041-8213/acdac6}{{\em Astrophys. J. Lett.}
  {\bfseries 951} no.~1, (2023) L8},
  \href{http://arxiv.org/abs/2306.16213}{{\ttfamily arXiv:2306.16213
  [astro-ph.HE]}}.

\bibitem{EPTA:2023fyk}
{\bfseries EPTA} Collaboration, J.~Antoniadis {\em et~al.}, ``{The second data
  release from the European Pulsar Timing Array III. Search for gravitational
  wave signals},'' \href{http://dx.doi.org/10.1051/0004-6361/202346844}{{\em
  Astron. Astrophys.} {\bfseries 678} (2023) A50},
  \href{http://arxiv.org/abs/2306.16214}{{\ttfamily arXiv:2306.16214
  [astro-ph.HE]}}.

\bibitem{Reardon:2023gzh}
D.~J. Reardon {\em et~al.}, ``{Search for an Isotropic Gravitational-wave
  Background with the Parkes Pulsar Timing Array},''
  \href{http://dx.doi.org/10.3847/2041-8213/acdd02}{{\em Astrophys. J. Lett.}
  {\bfseries 951} no.~1, (2023) L6},
  \href{http://arxiv.org/abs/2306.16215}{{\ttfamily arXiv:2306.16215
  [astro-ph.HE]}}.

\bibitem{Xu:2023wog}
H.~Xu {\em et~al.}, ``{Searching for the Nano-Hertz Stochastic Gravitational
  Wave Background with the Chinese Pulsar Timing Array Data Release I},''
  \href{http://dx.doi.org/10.1088/1674-4527/acdfa5}{{\em Res. Astron.
  Astrophys.} {\bfseries 23} no.~7, (2023) 075024},
  \href{http://arxiv.org/abs/2306.16216}{{\ttfamily arXiv:2306.16216
  [astro-ph.HE]}}.

\bibitem{Addazi:2023jvg}
A.~Addazi, Y.-F. Cai, A.~Marciano, and L.~Visinelli, ``{Have pulsar timing
  array methods detected a cosmological phase transition?},''
  \href{http://arxiv.org/abs/2306.17205}{{\ttfamily arXiv:2306.17205
  [astro-ph.CO]}}.

\bibitem{Kitajima:2023vre}
N.~Kitajima and K.~Nakayama, ``{Nanohertz gravitational waves from cosmic
  strings and dark photon dark matter},''
  \href{http://arxiv.org/abs/2306.17390}{{\ttfamily arXiv:2306.17390
  [hep-ph]}}.

\bibitem{Athron:2023mer}
P.~Athron, A.~Fowlie, C.-T. Lu, L.~Morris, L.~Wu, Y.~Wu, and Z.~Xu, ``{Can
  Supercooled Phase Transitions explain the Gravitational Wave Background
  observed by Pulsar Timing Arrays?},''
  \href{http://arxiv.org/abs/2306.17239}{{\ttfamily arXiv:2306.17239
  [hep-ph]}}.

\bibitem{Ghoshal:2023fhh}
A.~Ghoshal and A.~Strumia, ``{Probing the Dark Matter density with
  gravitational waves from super-massive binary black holes},''
  \href{http://arxiv.org/abs/2306.17158}{{\ttfamily arXiv:2306.17158
  [astro-ph.CO]}}.

\bibitem{Yang:2023aak}
J.~Yang, N.~Xie, and F.~P. Huang, ``{Nano-Hertz stochastic gravitational wave
  background as hints of ultralight axion particles},''
  \href{http://arxiv.org/abs/2306.17113}{{\ttfamily arXiv:2306.17113
  [hep-ph]}}.

\bibitem{Bai:2023cqj}
Y.~Bai, T.-K. Chen, and M.~Korwar, ``{QCD-Collapsed Domain Walls: QCD Phase
  Transition and Gravitational Wave Spectroscopy},''
  \href{http://arxiv.org/abs/2306.17160}{{\ttfamily arXiv:2306.17160
  [hep-ph]}}.

\bibitem{Ellis:2023dgf}
J.~Ellis, M.~Fairbairn, G.~H\"utsi, J.~Raidal, J.~Urrutia, V.~Vaskonen, and
  H.~Veerm\"ae, ``{Gravitational Waves from SMBH Binaries in Light of the
  NANOGrav 15-Year Data},'' \href{http://arxiv.org/abs/2306.17021}{{\ttfamily
  arXiv:2306.17021 [astro-ph.CO]}}.

\bibitem{Huang:2023chx}
H.-L. Huang, Y.~Cai, J.-Q. Jiang, J.~Zhang, and Y.-S. Piao, ``{Supermassive
  primordial black holes in multiverse: for nano-Hertz gravitational wave and
  high-redshift JWST galaxies},''
  \href{http://arxiv.org/abs/2306.17577}{{\ttfamily arXiv:2306.17577 [gr-qc]}}.

\bibitem{Franciolini:2023pbf}
G.~Franciolini, A.~Iovino, Junior., V.~Vaskonen, and H.~Veermae, ``{The recent
  gravitational wave observation by pulsar timing arrays and primordial black
  holes: the importance of non-gaussianities},''
  \href{http://arxiv.org/abs/2306.17149}{{\ttfamily arXiv:2306.17149
  [astro-ph.CO]}}.

\bibitem{Li:2023yaj}
Y.~Li, C.~Zhang, Z.~Wang, M.~Cui, Y.-L.~S. Tsai, Q.~Yuan, and Y.-Z. Fan,
  ``{Primordial magnetic field as a common solution of nanohertz gravitational
  waves and Hubble tension},''
  \href{http://arxiv.org/abs/2306.17124}{{\ttfamily arXiv:2306.17124
  [astro-ph.HE]}}.

\bibitem{Kitajima:2023cek}
N.~Kitajima, J.~Lee, K.~Murai, F.~Takahashi, and W.~Yin, ``{Nanohertz
  Gravitational Waves from Axion Domain Walls Coupled to QCD},''
  \href{http://arxiv.org/abs/2306.17146}{{\ttfamily arXiv:2306.17146
  [hep-ph]}}.

\bibitem{Datta:2023vbs}
S.~Datta and R.~Samanta, ``{Fingerprints of GeV scale right-handed neutrinos on
  inflationary gravitational waves and PTA data},''
  \href{http://dx.doi.org/10.1103/PhysRevD.108.L091706}{{\em Phys. Rev. D}
  {\bfseries 108} no.~9, (2023) L091706},
  \href{http://arxiv.org/abs/2307.00646}{{\ttfamily arXiv:2307.00646
  [hep-ph]}}.

\bibitem{Fujikura:2023lkn}
K.~Fujikura, S.~Girmohanta, Y.~Nakai, and M.~Suzuki, ``{NANOGrav Signal from a
  Dark Conformal Phase Transition},''
  \href{http://arxiv.org/abs/2306.17086}{{\ttfamily arXiv:2306.17086
  [hep-ph]}}.

\bibitem{Vagnozzi:2023lwo}
S.~Vagnozzi, ``{Inflationary interpretation of the stochastic gravitational
  wave background signal detected by pulsar timing array experiments},''
  \href{http://arxiv.org/abs/2306.16912}{{\ttfamily arXiv:2306.16912
  [astro-ph.CO]}}.

\bibitem{Ellis:2023tsl}
J.~Ellis, M.~Lewicki, C.~Lin, and V.~Vaskonen, ``{Cosmic Superstrings Revisited
  in Light of NANOGrav 15-Year Data},''
  \href{http://arxiv.org/abs/2306.17147}{{\ttfamily arXiv:2306.17147
  [astro-ph.CO]}}.

\bibitem{Wang:2023len}
Z.~Wang, L.~Lei, H.~Jiao, L.~Feng, and Y.-Z. Fan, ``{The nanohertz stochastic
  gravitational-wave background from cosmic string Loops and the abundant high
  redshift massive galaxies},''
  \href{http://arxiv.org/abs/2306.17150}{{\ttfamily arXiv:2306.17150
  [astro-ph.HE]}}.

\bibitem{Guo:2023hyp}
S.-Y. Guo, M.~Khlopov, X.~Liu, L.~Wu, Y.~Wu, and B.~Zhu, ``{Footprints of
  Axion-Like Particle in Pulsar Timing Array Data and JWST Observations},''
  \href{http://arxiv.org/abs/2306.17022}{{\ttfamily arXiv:2306.17022
  [hep-ph]}}.

\bibitem{Han:2023olf}
C.~Han, K.-P. Xie, J.~M. Yang, and M.~Zhang, ``{Self-interacting dark matter
  implied by nano-Hertz gravitational waves},''
  \href{http://arxiv.org/abs/2306.16966}{{\ttfamily arXiv:2306.16966
  [hep-ph]}}.

\bibitem{Shen:2023pan}
Z.-Q. Shen, G.-W. Yuan, Y.-Y. Wang, and Y.-Z. Wang, ``{Dark Matter Spike
  surrounding Supermassive Black Holes Binary and the nanohertz Stochastic
  Gravitational Wave Background},''
  \href{http://arxiv.org/abs/2306.17143}{{\ttfamily arXiv:2306.17143
  [astro-ph.HE]}}.

\bibitem{Zu:2023olm}
L.~Zu, C.~Zhang, Y.-Y. Li, Y.-C. Gu, Y.-L.~S. Tsai, and Y.-Z. Fan, ``{Mirror
  QCD phase transition as the origin of the nanohertz Stochastic
  Gravitational-Wave Background detected by the Pulsar Timing Arrays},''
  \href{http://arxiv.org/abs/2306.16769}{{\ttfamily arXiv:2306.16769
  [astro-ph.HE]}}.

\bibitem{Bi:2023tib}
Y.-C. Bi, Y.-M. Wu, Z.-C. Chen, and Q.-G. Huang, ``{Implications for the
  Supermassive Black Hole Binaries from the NANOGrav 15-year Data Set},''
  \href{http://arxiv.org/abs/2307.00722}{{\ttfamily arXiv:2307.00722
  [astro-ph.CO]}}.

\bibitem{Wang:2023ost}
S.~Wang, Z.-C. Zhao, J.-P. Li, and Q.-H. Zhu, ``{Exploring the Implications of
  2023 Pulsar Timing Array Datasets for Scalar-Induced Gravitational Waves and
  Primordial Black Holes},'' \href{http://arxiv.org/abs/2307.00572}{{\ttfamily
  arXiv:2307.00572 [astro-ph.CO]}}.

\bibitem{Broadhurst:2023tus}
T.~Broadhurst, C.~Chen, T.~Liu, and K.-F. Zheng, ``{Binary Supermassive Black
  Holes Orbiting Dark Matter Solitons: From the Dual AGN in UGC4211 to
  NanoHertz Gravitational Waves},''
  \href{http://arxiv.org/abs/2306.17821}{{\ttfamily arXiv:2306.17821
  [astro-ph.HE]}}.

\bibitem{Cai:2023dls}
Y.-F. Cai, X.-C. He, X.~Ma, S.-F. Yan, and G.-W. Yuan, ``{Limits on
  scalar-induced gravitational waves from the stochastic background by pulsar
  timing array observations},''
  \href{http://arxiv.org/abs/2306.17822}{{\ttfamily arXiv:2306.17822 [gr-qc]}}.

\bibitem{Depta:2023qst}
P.~F. Depta, K.~Schmidt-Hoberg, and C.~Tasillo, ``{Do pulsar timing arrays
  observe merging primordial black holes?},''
  \href{http://arxiv.org/abs/2306.17836}{{\ttfamily arXiv:2306.17836
  [astro-ph.CO]}}.

\bibitem{Bian:2023dnv}
L.~Bian, S.~Ge, J.~Shu, B.~Wang, X.-Y. Yang, and J.~Zong, ``{Gravitational wave
  sources for Pulsar Timing Arrays},''
  \href{http://arxiv.org/abs/2307.02376}{{\ttfamily arXiv:2307.02376
  [astro-ph.HE]}}.

\bibitem{Franciolini:2023wjm}
G.~Franciolini, D.~Racco, and F.~Rompineve, ``{Footprints of the QCD Crossover
  on Cosmological Gravitational Waves at Pulsar Timing Arrays},''
  \href{http://arxiv.org/abs/2306.17136}{{\ttfamily arXiv:2306.17136
  [astro-ph.CO]}}.

\bibitem{Du:2023qvj}
X.~K. Du, M.~X. Huang, F.~Wang, and Y.~K. Zhang, ``{Did the nHZ Gravitational
  Waves Signatures Observed By NANOGrav Indicate Multiple Sector SUSY
  Breaking?},'' \href{http://arxiv.org/abs/2307.02938}{{\ttfamily
  arXiv:2307.02938 [hep-ph]}}.

\bibitem{Wu:2023hsa}
Y.-M. Wu, Z.-C. Chen, and Q.-G. Huang, ``{Cosmological Interpretation for the
  Stochastic Signal in Pulsar Timing Arrays},''
  \href{http://arxiv.org/abs/2307.03141}{{\ttfamily arXiv:2307.03141
  [astro-ph.CO]}}.

\bibitem{Yi:2023mbm}
Z.~Yi, Q.~Gao, Y.~Gong, Y.~Wang, and F.~Zhang, ``{The waveform of the scalar
  induced gravitational waves in light of Pulsar Timing Array data},''
  \href{http://arxiv.org/abs/2307.02467}{{\ttfamily arXiv:2307.02467 [gr-qc]}}.

\bibitem{Liang:2023fdf}
Z.-C. Liang, Z.-Y. Li, E.-K. Li, J.-d. Zhang, and Y.-M. Hu, ``{Sensitivity to
  anisotropic stochastic gravitational-wave background with space-borne
  networks},'' \href{http://arxiv.org/abs/2307.01541}{{\ttfamily
  arXiv:2307.01541 [gr-qc]}}.

\bibitem{Anchordoqui:2023tln}
L.~A. Anchordoqui, I.~Antoniadis, and D.~Lust, ``{Fuzzy Dark Matter, the Dark
  Dimension, and the Pulsar Timing Array Signal},''
  \href{http://arxiv.org/abs/2307.01100}{{\ttfamily arXiv:2307.01100
  [hep-ph]}}.

\bibitem{Battista:2023znv}
E.~Battista, V.~De~Falco, and D.~Usseglio, ``{First post-Newtonian N-body
  problem in Einstein\textendash{}Cartan theory with the Weyssenhoff fluid:
  Lagrangian and first integrals},''
  \href{http://dx.doi.org/10.1140/epjc/s10052-023-11249-9}{{\em Eur. Phys. J.
  C} {\bfseries 83} no.~2, (2023) 112},
  \href{http://arxiv.org/abs/2301.08954}{{\ttfamily arXiv:2301.08954 [gr-qc]}}.

\bibitem{DeFalco:2023djo}
V.~De~Falco and E.~Battista, ``{Analytical results for binary dynamics at the
  first post-Newtonian order in Einstein-Cartan theory with the Weyssenhoff
  fluid},'' \href{http://dx.doi.org/10.1103/PhysRevD.108.064032}{{\em Phys.
  Rev. D} {\bfseries 108} no.~6, (2023) 064032},
  \href{http://arxiv.org/abs/2309.00319}{{\ttfamily arXiv:2309.00319 [gr-qc]}}.

\bibitem{Li:2023bxy}
S.-P. Li and K.-P. Xie, ``{Collider test of nano-Hertz gravitational waves from
  pulsar timing arrays},''
  \href{http://dx.doi.org/10.1103/PhysRevD.108.055018}{{\em Phys. Rev. D}
  {\bfseries 108} no.~5, (2023) 055018},
  \href{http://arxiv.org/abs/2307.01086}{{\ttfamily arXiv:2307.01086
  [hep-ph]}}.

\bibitem{Xiao:2023dbb}
Y.~Xiao, J.~M. Yang, and Y.~Zhang, ``{Implications of Nano-Hertz Gravitational
  Waves on Electroweak Phase Transition in the Singlet Dark Matter Model},''
  \href{http://arxiv.org/abs/2307.01072}{{\ttfamily arXiv:2307.01072
  [hep-ph]}}.

\bibitem{Zhang:2023lzt}
C.~Zhang, N.~Dai, Q.~Gao, Y.~Gong, T.~Jiang, and X.~Lu, ``{Detecting new
  fundamental fields with Pulsar Timing Arrays},''
  \href{http://arxiv.org/abs/2307.01093}{{\ttfamily arXiv:2307.01093 [gr-qc]}}.

\bibitem{Liu:2023ymk}
L.~Liu, Z.-C. Chen, and Q.-G. Huang, ``{Implications for the non-Gaussianity of
  curvature perturbation from pulsar timing arrays},''
  \href{http://arxiv.org/abs/2307.01102}{{\ttfamily arXiv:2307.01102
  [astro-ph.CO]}}.

\bibitem{Inomata:2023zup}
K.~Inomata, K.~Kohri, and T.~Terada, ``{The Detected Stochastic Gravitational
  Waves and Subsolar-Mass Primordial Black Holes},''
  \href{http://arxiv.org/abs/2306.17834}{{\ttfamily arXiv:2306.17834
  [astro-ph.CO]}}.

\bibitem{Ghosh:2023aum}
T.~Ghosh, A.~Ghoshal, H.-K. Guo, F.~Hajkarim, S.~F. King, K.~Sinha, X.~Wang,
  and G.~White, ``{Did we hear the sound of the Universe boiling? Analysis
  using the full fluid velocity profiles and NANOGrav 15-year data},''
  \href{http://arxiv.org/abs/2307.02259}{{\ttfamily arXiv:2307.02259
  [astro-ph.HE]}}.

\bibitem{Niu:2023bsr}
X.~Niu and M.~H. Rahat, ``{NANOGrav signal from axion inflation},''
  \href{http://arxiv.org/abs/2307.01192}{{\ttfamily arXiv:2307.01192
  [hep-ph]}}.

\bibitem{Konoplya:2023fmh}
R.~A. Konoplya and A.~Zhidenko, ``{Asymptotic tails of massive gravitons in
  light of pulsar timing array observations},''
  \href{http://arxiv.org/abs/2307.01110}{{\ttfamily arXiv:2307.01110 [gr-qc]}}.

\bibitem{DiBari:2023upq}
P.~Di~Bari and M.~H. Rahat, ``{The split majoron model confronts the NANOGrav
  signal},'' \href{http://arxiv.org/abs/2307.03184}{{\ttfamily arXiv:2307.03184
  [hep-ph]}}.

\bibitem{Wang:2023sij}
S.~Wang, Z.-C. Zhao, and Q.-H. Zhu, ``{Constraints On Scalar-Induced
  Gravitational Waves Up To Third Order From Joint Analysis of BBN, CMB, And
  PTA Data},'' \href{http://arxiv.org/abs/2307.03095}{{\ttfamily
  arXiv:2307.03095 [astro-ph.CO]}}.

\bibitem{Ye:2023xyr}
G.~Ye and A.~Silvestri, ``{Can the gravitational wave background feel wiggles
  in spacetime?},'' \href{http://arxiv.org/abs/2307.05455}{{\ttfamily
  arXiv:2307.05455 [astro-ph.CO]}}.

\bibitem{Balaji:2023ehk}
S.~Balaji, G.~Dom\`enech, and G.~Franciolini, ``{Scalar-induced gravitational
  wave interpretation of PTA data: the role of scalar fluctuation propagation
  speed},'' \href{http://arxiv.org/abs/2307.08552}{{\ttfamily arXiv:2307.08552
  [gr-qc]}}.

\bibitem{Zhang:2023nrs}
Z.~Zhang, C.~Cai, Y.-H. Su, S.~Wang, Z.-H. Yu, and H.-H. Zhang, ``{Nano-Hertz
  gravitational waves from collapsing domain walls associated with freeze-in
  dark matter in light of pulsar timing array observations},''
  \href{http://arxiv.org/abs/2307.11495}{{\ttfamily arXiv:2307.11495
  [hep-ph]}}.

\bibitem{Jiang:2023gfe}
J.-Q. Jiang, Y.~Cai, G.~Ye, and Y.-S. Piao, ``{Broken blue-tilted inflationary
  gravitational waves: a joint analysis of NANOGrav 15-year and BICEP/Keck 2018
  data},'' \href{http://arxiv.org/abs/2307.15547}{{\ttfamily arXiv:2307.15547
  [astro-ph.CO]}}.

\bibitem{Zhu:2023lbf}
M.~Zhu, G.~Ye, and Y.~Cai, ``{Pulsar timing array observations as possible
  hints for nonsingular cosmology},''
  \href{http://dx.doi.org/10.1140/epjc/s10052-023-11963-4}{{\em Eur. Phys. J.
  C} {\bfseries 83} no.~9, (2023) 816},
  \href{http://arxiv.org/abs/2307.16211}{{\ttfamily arXiv:2307.16211
  [astro-ph.CO]}}.

\bibitem{An:2023jxf}
H.~An, B.~Su, H.~Tai, L.-T. Wang, and C.~Yang, ``{Phase transition during
  inflation and the gravitational wave signal at pulsar timing arrays},''
  \href{http://arxiv.org/abs/2308.00070}{{\ttfamily arXiv:2308.00070
  [astro-ph.CO]}}.

\bibitem{Hooper:2023nnl}
D.~Hooper, A.~Ireland, G.~Krnjaic, and A.~Stebbins, ``{Supermassive Primordial
  Black Holes From Inflation},''
  \href{http://arxiv.org/abs/2308.00756}{{\ttfamily arXiv:2308.00756
  [astro-ph.CO]}}.

\bibitem{King:2023ayw}
S.~F. King, R.~Roshan, X.~Wang, G.~White, and M.~Yamazaki, ``{Quantum Gravity
  Effects on Dark Matter and Gravitational Waves},''
  \href{http://arxiv.org/abs/2308.03724}{{\ttfamily arXiv:2308.03724
  [hep-ph]}}.

\bibitem{Maji:2023fhv}
R.~Maji and W.-I. Park, ``{Supersymmetric $U(1)_{B-L}$ flat direction and
  NANOGrav 15 year data},'' \href{http://arxiv.org/abs/2308.11439}{{\ttfamily
  arXiv:2308.11439 [hep-ph]}}.

\bibitem{Datta:2023xpr}
S.~Datta, ``{Explaining PTA Data with Inflationary GWs in a PBH-Dominated
  Universe},'' \href{http://arxiv.org/abs/2309.14238}{{\ttfamily
  arXiv:2309.14238 [hep-ph]}}.

\bibitem{HosseiniMansoori:2023mqh}
S.~A. Hosseini~Mansoori, F.~Felegray, A.~Talebian, and M.~Sami, ``{PBHs and GWs
  from \ensuremath{\mathbb{T}}$^{2}$-inflation and NANOGrav 15-year data},''
  \href{http://dx.doi.org/10.1088/1475-7516/2023/08/067}{{\em JCAP} {\bfseries
  08} (2023) 067}, \href{http://arxiv.org/abs/2307.06757}{{\ttfamily
  arXiv:2307.06757 [astro-ph.CO]}}.

\bibitem{Frosina:2023nxu}
L.~Frosina and A.~Urbano, ``{On the inflationary interpretation of the nHz
  gravitational-wave background},''
  \href{http://arxiv.org/abs/2308.06915}{{\ttfamily arXiv:2308.06915
  [astro-ph.CO]}}.

\bibitem{He:2023ado}
S.~He, L.~Li, S.~Wang, and S.-J. Wang, ``{Constraints on holographic QCD phase
  transitions from PTA observations},''
  \href{http://arxiv.org/abs/2308.07257}{{\ttfamily arXiv:2308.07257
  [hep-ph]}}.

\bibitem{Ellis:2023oxs}
J.~Ellis, M.~Fairbairn, G.~Franciolini, G.~H\"utsi, A.~Iovino, M.~Lewicki,
  M.~Raidal, J.~Urrutia, V.~Vaskonen, and H.~Veerm\"ae, ``{What is the source
  of the PTA GW signal?},'' \href{http://arxiv.org/abs/2308.08546}{{\ttfamily
  arXiv:2308.08546 [astro-ph.CO]}}.

\bibitem{Kawasaki:2023rfx}
M.~Kawasaki and K.~Murai, ``{Enhancement of gravitational waves at Q-ball decay
  including non-linear density perturbations},''
  \href{http://arxiv.org/abs/2308.13134}{{\ttfamily arXiv:2308.13134
  [astro-ph.CO]}}.

\bibitem{Kawai:2023nqs}
S.~Kawai and J.~Kim, ``{Probing inflationary moduli space with gravitational
  waves},'' \href{http://arxiv.org/abs/2308.13272}{{\ttfamily arXiv:2308.13272
  [astro-ph.CO]}}.

\bibitem{Bhattacharya:2023ysp}
G.~Bhattacharya, S.~Choudhury, K.~Dey, S.~Ghosh, A.~Karde, and N.~S. Mishra,
  ``{Evading no-go for PBH formation and production of SIGWs using Multiple
  Sharp Transitions in EFT of single field inflation},''
  \href{http://arxiv.org/abs/2309.00973}{{\ttfamily arXiv:2309.00973
  [astro-ph.CO]}}.

\bibitem{Huang:2023zvs}
M.~X. Huang, F.~Wang, and Y.~K. Zhang, ``{The Interplay Between the Muon $g-2$
  Anomaly and the PTA nHZ Gravitational Waves from Domain Walls in NMSSM},''
  \href{http://arxiv.org/abs/2309.06378}{{\ttfamily arXiv:2309.06378
  [hep-ph]}}.

\bibitem{Lozanov:2023rcd}
K.~D. Lozanov, S.~Pi, M.~Sasaki, V.~Takhistov, and A.~Wang, ``{Axion Universal
  Gravitational Wave Interpretation of Pulsar Timing Array Data},''
  \href{http://arxiv.org/abs/2310.03594}{{\ttfamily arXiv:2310.03594
  [astro-ph.CO]}}.

\bibitem{Bernardo:2023zna}
R.~C. Bernardo and K.-W. Ng, ``{Beyond the Hellings-Downs curve:
  Non-Einsteinian gravitational waves in pulsar timing array correlations},''
  \href{http://arxiv.org/abs/2310.07537}{{\ttfamily arXiv:2310.07537 [gr-qc]}}.

\bibitem{Chen:2023bms}
Z.-C. Chen, S.-L. Li, P.~Wu, and H.~Yu, ``{NANOGrav hints for first-order
  confinement-deconfinement phase transition in different QCD-matter
  scenarios},'' \href{http://arxiv.org/abs/2312.01824}{{\ttfamily
  arXiv:2312.01824 [astro-ph.CO]}}.

\bibitem{Ahmadvand:2023lpp}
M.~Ahmadvand, L.~Bian, and S.~Shakeri, ``{Heavy QCD axion model in light of
  pulsar timing arrays},''
  \href{http://dx.doi.org/10.1103/PhysRevD.108.115020}{{\em Phys. Rev. D}
  {\bfseries 108} no.~11, (2023) 115020},
  \href{http://arxiv.org/abs/2307.12385}{{\ttfamily arXiv:2307.12385
  [hep-ph]}}.

\bibitem{Chowdhury:2023xvy}
D.~Chowdhury, A.~Hait, S.~Mohanty, and S.~Prakash, ``{Ultralight
  $(L_\mu-L_\tau)$ vector dark matter interpretation of NANOGrav
  observations},'' \href{http://arxiv.org/abs/2311.10148}{{\ttfamily
  arXiv:2311.10148 [hep-ph]}}.

\bibitem{Aghaie:2023lan}
M.~Aghaie, G.~Armando, A.~Dondarini, and P.~Panci, ``{Bounds on Ultralight Dark
  Matter from NANOGrav},'' \href{http://arxiv.org/abs/2308.04590}{{\ttfamily
  arXiv:2308.04590 [astro-ph.CO]}}.

\bibitem{Choudhury:2024one}
S.~Choudhury, A.~Karde, S.~Panda, and M.~Sami, ``{Realisation of the ultra-slow
  roll phase in Galileon inflation and PBH overproduction},''
  \href{http://arxiv.org/abs/2401.10925}{{\ttfamily arXiv:2401.10925
  [astro-ph.CO]}}.

\bibitem{Choudhury:2023fjs}
S.~Choudhury, K.~Dey, and A.~Karde, ``{Untangling PBH overproduction in
  $w$-SIGWs generated by Pulsar Timing Arrays for MST-EFT of single field
  inflation},'' \href{http://arxiv.org/abs/2311.15065}{{\ttfamily
  arXiv:2311.15065 [astro-ph.CO]}}.

\bibitem{Choudhury:2023fwk}
S.~Choudhury, K.~Dey, A.~Karde, S.~Panda, and M.~Sami, ``{Primordial
  non-Gaussianity as a saviour for PBH overproduction in SIGWs generated by
  Pulsar Timing Arrays for Galileon inflation},''
  \href{http://arxiv.org/abs/2310.11034}{{\ttfamily arXiv:2310.11034
  [astro-ph.CO]}}.

\bibitem{Choudhury:2023hfm}
S.~Choudhury, A.~Karde, S.~Panda, and M.~Sami, ``{Scalar induced gravity waves
  from ultra slow-roll Galileon inflation},''
  \href{http://arxiv.org/abs/2308.09273}{{\ttfamily arXiv:2308.09273
  [astro-ph.CO]}}.

\bibitem{Choudhury:2013woa}
S.~Choudhury and A.~Mazumdar, ``{Primordial blackholes and gravitational waves
  for an inflection-point model of inflation},''
  \href{http://dx.doi.org/10.1016/j.physletb.2014.04.050}{{\em Phys. Lett. B}
  {\bfseries 733} (2014) 270--275},
  \href{http://arxiv.org/abs/1307.5119}{{\ttfamily arXiv:1307.5119
  [astro-ph.CO]}}.

\bibitem{Grishchuk:1974ny}
L.~P. Grishchuk, ``{Amplification of gravitational waves in an istropic
  universe},'' {\em Zh. Eksp. Teor. Fiz.} {\bfseries 67} (1974) 825--838.

\bibitem{Starobinsky:1979ty}
A.~A. Starobinsky, ``{Spectrum of relict gravitational radiation and the early
  state of the universe},'' {\em JETP Lett.} {\bfseries 30} (1979) 682--685.

\bibitem{Rubakov:1982df}
V.~A. Rubakov, M.~V. Sazhin, and A.~V. Veryaskin, ``{Graviton Creation in the
  Inflationary Universe and the Grand Unification Scale},''
  \href{http://dx.doi.org/10.1016/0370-2693(82)90641-4}{{\em Phys. Lett. B}
  {\bfseries 115} (1982) 189--192}.

\bibitem{Borah:2023sbc}
D.~Borah, S.~Jyoti~Das, and R.~Samanta, ``{Inflationary origin of gravitational
  waves with \textbackslash{}textit{Miracle-less WIMP} dark matter in the light
  of recent PTA results},'' \href{http://arxiv.org/abs/2307.00537}{{\ttfamily
  arXiv:2307.00537 [hep-ph]}}.

\bibitem{Choudhury:2023kam}
S.~Choudhury, ``{Single field inflation in the light of NANOGrav 15-year Data:
  Quintessential interpretation of blue tilted tensor spectrum through
  Non-Bunch Davies initial condition},''
  \href{http://arxiv.org/abs/2307.03249}{{\ttfamily arXiv:2307.03249
  [astro-ph.CO]}}.

\bibitem{Ben-Dayan:2023lwd}
I.~Ben-Dayan, U.~Kumar, U.~Thattarampilly, and A.~Verma, ``{Probing The Early
  Universe Cosmology With NANOGrav: Possibilities and Limitations},''
  \href{http://arxiv.org/abs/2307.15123}{{\ttfamily arXiv:2307.15123
  [astro-ph.CO]}}.

\bibitem{Oikonomou:2023qfz}
V.~K. Oikonomou, ``{Flat energy spectrum of primordial gravitational waves
  versus peaks and the NANOGrav 2023 observation},''
  \href{http://dx.doi.org/10.1103/PhysRevD.108.043516}{{\em Phys. Rev. D}
  {\bfseries 108} no.~4, (2023) 043516},
  \href{http://arxiv.org/abs/2306.17351}{{\ttfamily arXiv:2306.17351
  [astro-ph.CO]}}.

\bibitem{Guth:1980zm}
A.~H. Guth, ``{The Inflationary Universe: A Possible Solution to the Horizon
  and Flatness Problems},''
  \href{http://dx.doi.org/10.1103/PhysRevD.23.347}{{\em Phys. Rev.} {\bfseries
  D23} (1981) 347--356}.
[Adv. Ser. Astrophys. Cosmol.3,139(1987)].
%%CITATION = PHRVA,D23,347;%%.

\bibitem{Linde:1981mu}
A.~D. Linde, ``{A New Inflationary Universe Scenario: A Possible Solution of
  the Horizon, Flatness, Homogeneity, Isotropy and Primordial Monopole
  Problems},'' \href{http://dx.doi.org/10.1016/0370-2693(82)91219-9}{{\em Phys.
  Lett.} {\bfseries 108B} (1982) 389--393}.
[Adv. Ser. Astrophys. Cosmol.3,149(1987)].
%%CITATION = PHLTA,108B,389;%%.

\bibitem{Albrecht:1982wi}
A.~Albrecht and P.~J. Steinhardt, ``{Cosmology for Grand Unified Theories with
  Radiatively Induced Symmetry Breaking},''
  \href{http://dx.doi.org/10.1103/PhysRevLett.48.1220}{{\em Phys. Rev. Lett.}
  {\bfseries 48} (1982) 1220--1223}.
[Adv. Ser. Astrophys. Cosmol.3,158(1987)].
%%CITATION = PRLTA,48,1220;%%.

\bibitem{Starobinsky:1980te}
A.~A. Starobinsky, ``{A New Type of Isotropic Cosmological Models Without
  Singularity},'' \href{http://dx.doi.org/10.1016/0370-2693(80)90670-X}{{\em
  Phys. Lett. B} {\bfseries 91} (1980) 99--102}.

\bibitem{Piao:2004tq}
Y.-S. Piao and Y.-Z. Zhang, ``{Phantom inflation and primordial perturbation
  spectrum},'' \href{http://dx.doi.org/10.1103/PhysRevD.70.063513}{{\em Phys.
  Rev. D} {\bfseries 70} (2004) 063513},
  \href{http://arxiv.org/abs/astro-ph/0401231}{{\ttfamily
  arXiv:astro-ph/0401231}}.

\bibitem{Baldi:2005gk}
M.~Baldi, F.~Finelli, and S.~Matarrese, ``{Inflation with violation of the null
  energy condition},'' \href{http://dx.doi.org/10.1103/PhysRevD.72.083504}{{\em
  Phys. Rev. D} {\bfseries 72} (2005) 083504},
  \href{http://arxiv.org/abs/astro-ph/0505552}{{\ttfamily
  arXiv:astro-ph/0505552}}.

\bibitem{Piao:2006jz}
Y.-S. Piao, ``{Gravitational wave background from phantom superinflation},''
  \href{http://dx.doi.org/10.1103/PhysRevD.73.047302}{{\em Phys. Rev. D}
  {\bfseries 73} (2006) 047302},
  \href{http://arxiv.org/abs/gr-qc/0601115}{{\ttfamily arXiv:gr-qc/0601115}}.

\bibitem{Kobayashi:2010cm}
T.~Kobayashi, M.~Yamaguchi, and J.~Yokoyama, ``{G-inflation: Inflation driven
  by the Galileon field},''
  \href{http://dx.doi.org/10.1103/PhysRevLett.105.231302}{{\em Phys. Rev.
  Lett.} {\bfseries 105} (2010) 231302},
  \href{http://arxiv.org/abs/1008.0603}{{\ttfamily arXiv:1008.0603 [hep-th]}}.

\bibitem{Kobayashi:2011nu}
T.~Kobayashi, M.~Yamaguchi, and J.~Yokoyama, ``{Generalized G-inflation:
  Inflation with the most general second-order field equations},''
  \href{http://dx.doi.org/10.1143/PTP.126.511}{{\em Prog. Theor. Phys.}
  {\bfseries 126} (2011) 511--529},
  \href{http://arxiv.org/abs/1105.5723}{{\ttfamily arXiv:1105.5723 [hep-th]}}.

\bibitem{Endlich:2012pz}
S.~Endlich, A.~Nicolis, and J.~Wang, ``{Solid Inflation},''
  \href{http://dx.doi.org/10.1088/1475-7516/2013/10/011}{{\em JCAP} {\bfseries
  10} (2013) 011}, \href{http://arxiv.org/abs/1210.0569}{{\ttfamily
  arXiv:1210.0569 [hep-th]}}.

\bibitem{Cai:2014uka}
Y.-F. Cai, J.-O. Gong, S.~Pi, E.~N. Saridakis, and S.-Y. Wu, ``{On the
  possibility of blue tensor spectrum within single field inflation},''
  \href{http://dx.doi.org/10.1016/j.nuclphysb.2015.09.025}{{\em Nucl. Phys. B}
  {\bfseries 900} (2015) 517--532},
  \href{http://arxiv.org/abs/1412.7241}{{\ttfamily arXiv:1412.7241 [hep-th]}}.

\bibitem{Gong:2014qga}
J.-O. Gong, ``{Blue running of the primordial tensor spectrum},''
  \href{http://dx.doi.org/10.1088/1475-7516/2014/07/022}{{\em JCAP} {\bfseries
  07} (2014) 022}, \href{http://arxiv.org/abs/1403.5163}{{\ttfamily
  arXiv:1403.5163 [astro-ph.CO]}}.

\bibitem{Cannone:2014uqa}
D.~Cannone, G.~Tasinato, and D.~Wands, ``{Generalised tensor fluctuations and
  inflation},'' \href{http://dx.doi.org/10.1088/1475-7516/2015/01/029}{{\em
  JCAP} {\bfseries 01} (2015) 029},
  \href{http://arxiv.org/abs/1409.6568}{{\ttfamily arXiv:1409.6568
  [astro-ph.CO]}}.

\bibitem{Wang:2014kqa}
Y.~Wang and W.~Xue, ``{Inflation and Alternatives with Blue Tensor Spectra},''
  \href{http://dx.doi.org/10.1088/1475-7516/2014/10/075}{{\em JCAP} {\bfseries
  10} (2014) 075}, \href{http://arxiv.org/abs/1403.5817}{{\ttfamily
  arXiv:1403.5817 [astro-ph.CO]}}.

\bibitem{Kuroyanagi:2014nba}
S.~Kuroyanagi, T.~Takahashi, and S.~Yokoyama, ``{Blue-tilted Tensor Spectrum
  and Thermal History of the Universe},''
  \href{http://dx.doi.org/10.1088/1475-7516/2015/02/003}{{\em JCAP} {\bfseries
  02} (2015) 003}, \href{http://arxiv.org/abs/1407.4785}{{\ttfamily
  arXiv:1407.4785 [astro-ph.CO]}}.

\bibitem{Cai:2015yza}
Y.~Cai, Y.-T. Wang, and Y.-S. Piao, ``{Is there an effect of a nontrivial $c_T$
  during inflation?},''
  \href{http://dx.doi.org/10.1103/PhysRevD.93.063005}{{\em Phys. Rev. D}
  {\bfseries 93} no.~6, (2016) 063005},
  \href{http://arxiv.org/abs/1510.08716}{{\ttfamily arXiv:1510.08716
  [astro-ph.CO]}}.

\bibitem{Cai:2016ldn}
Y.~Cai, Y.-T. Wang, and Y.-S. Piao, ``{Propagating speed of primordial
  gravitational waves and inflation},''
  \href{http://dx.doi.org/10.1103/PhysRevD.94.043002}{{\em Phys. Rev. D}
  {\bfseries 94} no.~4, (2016) 043002},
  \href{http://arxiv.org/abs/1602.05431}{{\ttfamily arXiv:1602.05431
  [astro-ph.CO]}}.

\bibitem{Wang:2016tbj}
Y.-T. Wang, Y.~Cai, Z.-G. Liu, and Y.-S. Piao, ``{Probing the primordial
  universe with gravitational waves detectors},''
  \href{http://dx.doi.org/10.1088/1475-7516/2017/01/010}{{\em JCAP} {\bfseries
  01} (2017) 010}, \href{http://arxiv.org/abs/1612.05088}{{\ttfamily
  arXiv:1612.05088 [astro-ph.CO]}}.

\bibitem{Fujita:2018ehq}
T.~Fujita, S.~Kuroyanagi, S.~Mizuno, and S.~Mukohyama, ``{Blue-tilted
  Primordial Gravitational Waves from Massive Gravity},''
  \href{http://dx.doi.org/10.1016/j.physletb.2018.12.025}{{\em Phys. Lett. B}
  {\bfseries 789} (2019) 215--219},
  \href{http://arxiv.org/abs/1808.02381}{{\ttfamily arXiv:1808.02381 [gr-qc]}}.

\bibitem{Kuroyanagi:2020sfw}
S.~Kuroyanagi, T.~Takahashi, and S.~Yokoyama, ``{Blue-tilted inflationary
  tensor spectrum and reheating in the light of NANOGrav results},''
  \href{http://dx.doi.org/10.1088/1475-7516/2021/01/071}{{\em JCAP} {\bfseries
  01} (2021) 071}, \href{http://arxiv.org/abs/2011.03323}{{\ttfamily
  arXiv:2011.03323 [astro-ph.CO]}}.

\bibitem{Akama:2020jko}
S.~Akama, S.~Hirano, and T.~Kobayashi, ``{Primordial tensor non-Gaussianities
  from general single-field inflation with non-Bunch-Davies initial states},''
  \href{http://dx.doi.org/10.1103/PhysRevD.102.023513}{{\em Phys. Rev. D}
  {\bfseries 102} no.~2, (2020) 023513},
  \href{http://arxiv.org/abs/2003.10686}{{\ttfamily arXiv:2003.10686 [gr-qc]}}.

\bibitem{Akama:2023jsb}
S.~Akama and H.~W.~H. Tahara, ``{Imprints of primordial gravitational waves
  with non-Bunch-Davies initial states on CMB bispectra},''
  \href{http://arxiv.org/abs/2306.17752}{{\ttfamily arXiv:2306.17752 [gr-qc]}}.

\bibitem{Giare:2020plo}
W.~Giar\`e, F.~Renzi, and A.~Melchiorri, ``{Higher-Curvature Corrections and
  Tensor Modes},'' \href{http://dx.doi.org/10.1103/PhysRevD.103.043515}{{\em
  Phys. Rev. D} {\bfseries 103} no.~4, (2021) 043515},
  \href{http://arxiv.org/abs/2012.00527}{{\ttfamily arXiv:2012.00527
  [astro-ph.CO]}}.

\bibitem{Giare:2022wxq}
W.~Giar\`e, M.~Forconi, E.~Di~Valentino, and A.~Melchiorri, ``{Towards a
  reliable calculation of relic radiation from primordial gravitational
  waves},'' \href{http://dx.doi.org/10.1093/mnras/stad258}{{\em Mon. Not. Roy.
  Astron. Soc.} {\bfseries 520} (2023) 2},
  \href{http://arxiv.org/abs/2210.14159}{{\ttfamily arXiv:2210.14159
  [astro-ph.CO]}}.

\bibitem{Rubakov:2014jja}
V.~A. Rubakov, ``{The Null Energy Condition and its violation},''
  \href{http://dx.doi.org/10.3367/UFNe.0184.201402b.0137}{{\em Phys. Usp.}
  {\bfseries 57} (2014) 128--142},
  \href{http://arxiv.org/abs/1401.4024}{{\ttfamily arXiv:1401.4024 [hep-th]}}.

\bibitem{Cai:2016thi}
Y.~Cai, Y.~Wan, H.-G. Li, T.~Qiu, and Y.-S. Piao, ``{The Effective Field Theory
  of nonsingular cosmology},''
  \href{http://dx.doi.org/10.1007/JHEP01(2017)090}{{\em JHEP} {\bfseries 01}
  (2017) 090}, \href{http://arxiv.org/abs/1610.03400}{{\ttfamily
  arXiv:1610.03400 [gr-qc]}}.

\bibitem{Creminelli:2016zwa}
P.~Creminelli, D.~Pirtskhalava, L.~Santoni, and E.~Trincherini, ``{Stability of
  Geodesically Complete Cosmologies},''
  \href{http://dx.doi.org/10.1088/1475-7516/2016/11/047}{{\em JCAP} {\bfseries
  11} (2016) 047}, \href{http://arxiv.org/abs/1610.04207}{{\ttfamily
  arXiv:1610.04207 [hep-th]}}.

\bibitem{Cai:2017tku}
Y.~Cai, H.-G. Li, T.~Qiu, and Y.-S. Piao, ``{The Effective Field Theory of
  nonsingular cosmology: II},''
  \href{http://dx.doi.org/10.1140/epjc/s10052-017-4938-y}{{\em Eur. Phys. J. C}
  {\bfseries 77} no.~6, (2017) 369},
  \href{http://arxiv.org/abs/1701.04330}{{\ttfamily arXiv:1701.04330 [gr-qc]}}.

\bibitem{Cai:2017dyi}
Y.~Cai and Y.-S. Piao, ``{A covariant Lagrangian for stable nonsingular
  bounce},'' \href{http://dx.doi.org/10.1007/JHEP09(2017)027}{{\em JHEP}
  {\bfseries 09} (2017) 027}, \href{http://arxiv.org/abs/1705.03401}{{\ttfamily
  arXiv:1705.03401 [gr-qc]}}.

\bibitem{Kolevatov:2017voe}
R.~Kolevatov, S.~Mironov, N.~Sukhov, and V.~Volkova, ``{Cosmological bounce and
  Genesis beyond Horndeski},''
  \href{http://dx.doi.org/10.1088/1475-7516/2017/08/038}{{\em JCAP} {\bfseries
  08} (2017) 038}, \href{http://arxiv.org/abs/1705.06626}{{\ttfamily
  arXiv:1705.06626 [hep-th]}}.

\bibitem{Ilyas:2020zcb}
A.~Ilyas, M.~Zhu, Y.~Zheng, and Y.-F. Cai, ``{Emergent Universe and Genesis
  from the DHOST Cosmology},''
  \href{http://dx.doi.org/10.1007/JHEP01(2021)141}{{\em JHEP} {\bfseries 01}
  (2021) 141}, \href{http://arxiv.org/abs/2009.10351}{{\ttfamily
  arXiv:2009.10351 [gr-qc]}}.

\bibitem{Ilyas:2020qja}
A.~Ilyas, M.~Zhu, Y.~Zheng, Y.-F. Cai, and E.~N. Saridakis, ``{DHOST Bounce},''
  \href{http://dx.doi.org/10.1088/1475-7516/2020/09/002}{{\em JCAP} {\bfseries
  09} (2020) 002}, \href{http://arxiv.org/abs/2002.08269}{{\ttfamily
  arXiv:2002.08269 [gr-qc]}}.

\bibitem{Zhu:2021ggm}
M.~Zhu and Y.~Zheng, ``{Improved DHOST Genesis},''
  \href{http://dx.doi.org/10.1007/JHEP11(2021)163}{{\em JHEP} {\bfseries 11}
  (2021) 163}, \href{http://arxiv.org/abs/2109.05277}{{\ttfamily
  arXiv:2109.05277 [gr-qc]}}.

\bibitem{Libanov:2016kfc}
M.~Libanov, S.~Mironov, and V.~Rubakov, ``{Generalized Galileons: instabilities
  of bouncing and Genesis cosmologies and modified Genesis},''
  \href{http://dx.doi.org/10.1088/1475-7516/2016/08/037}{{\em JCAP} {\bfseries
  08} (2016) 037}, \href{http://arxiv.org/abs/1605.05992}{{\ttfamily
  arXiv:1605.05992 [hep-th]}}.

\bibitem{Kobayashi:2016xpl}
T.~Kobayashi, ``{Generic instabilities of nonsingular cosmologies in Horndeski
  theory: A no-go theorem},''
  \href{http://dx.doi.org/10.1103/PhysRevD.94.043511}{{\em Phys. Rev. D}
  {\bfseries 94} no.~4, (2016) 043511},
  \href{http://arxiv.org/abs/1606.05831}{{\ttfamily arXiv:1606.05831
  [hep-th]}}.

\bibitem{Ijjas:2016vtq}
A.~Ijjas and P.~J. Steinhardt, ``{Fully stable cosmological solutions with a
  non-singular classical bounce},''
  \href{http://dx.doi.org/10.1016/j.physletb.2016.11.047}{{\em Phys. Lett. B}
  {\bfseries 764} (2017) 289--294},
  \href{http://arxiv.org/abs/1609.01253}{{\ttfamily arXiv:1609.01253 [gr-qc]}}.

\bibitem{Dobre:2017pnt}
D.~A. Dobre, A.~V. Frolov, J.~T.~G. Ghersi, S.~Ramazanov, and A.~Vikman,
  ``{Unbraiding the Bounce: Superluminality around the Corner},''
  \href{http://dx.doi.org/10.1088/1475-7516/2018/03/020}{{\em JCAP} {\bfseries
  03} (2018) 020}, \href{http://arxiv.org/abs/1712.10272}{{\ttfamily
  arXiv:1712.10272 [gr-qc]}}.

\bibitem{Zhu:2021whu}
M.~Zhu, A.~Ilyas, Y.~Zheng, Y.-F. Cai, and E.~N. Saridakis, ``{Scalar and
  tensor perturbations in DHOST bounce cosmology},''
  \href{http://dx.doi.org/10.1088/1475-7516/2021/11/045}{{\em JCAP} {\bfseries
  11} no.~11, (2021) 045}, \href{http://arxiv.org/abs/2108.01339}{{\ttfamily
  arXiv:2108.01339 [gr-qc]}}.

\bibitem{Cai:2022ori}
Y.~Cai, J.~Xu, S.~Zhao, and S.~Zhou, ``{Perturbative unitarity and NEC
  violation in genesis cosmology},''
  \href{http://dx.doi.org/10.1007/JHEP10(2022)140}{{\em JHEP} {\bfseries 10}
  (2022) 140}, \href{http://arxiv.org/abs/2207.11772}{{\ttfamily
  arXiv:2207.11772 [gr-qc]}}. [Erratum: JHEP 11, 063 (2022)].

\bibitem{Cai:2020qpu}
Y.~Cai and Y.-S. Piao, ``{Intermittent null energy condition violations during
  inflation and primordial gravitational waves},''
  \href{http://dx.doi.org/10.1103/PhysRevD.103.083521}{{\em Phys. Rev. D}
  {\bfseries 103} no.~8, (2021) 083521},
  \href{http://arxiv.org/abs/2012.11304}{{\ttfamily arXiv:2012.11304 [gr-qc]}}.

\bibitem{Cai:2022nqv}
Y.~Cai and Y.-S. Piao, ``{Generating enhanced primordial GWs during inflation
  with intermittent violation of NEC and diminishment of GW propagating
  speed},'' \href{http://dx.doi.org/10.1007/JHEP06(2022)067}{{\em JHEP}
  {\bfseries 06} (2022) 067}, \href{http://arxiv.org/abs/2201.04552}{{\ttfamily
  arXiv:2201.04552 [gr-qc]}}.

\bibitem{Cai:2022lec}
Y.~Cai, ``{Generating enhanced parity-violating gravitational waves during
  inflation with violation of the null energy condition},''
  \href{http://dx.doi.org/10.1103/PhysRevD.107.063512}{{\em Phys. Rev. D}
  {\bfseries 107} no.~6, (2023) 063512},
  \href{http://arxiv.org/abs/2212.10893}{{\ttfamily arXiv:2212.10893 [gr-qc]}}.

\bibitem{Zhu:2023lhv}
M.~Zhu and Y.~Cai, ``{Parity-violation in bouncing cosmology},''
  \href{http://dx.doi.org/10.1007/JHEP04(2023)095}{{\em JHEP} {\bfseries 04}
  (2023) 095}, \href{http://arxiv.org/abs/2301.13502}{{\ttfamily
  arXiv:2301.13502 [gr-qc]}}.

\bibitem{Cai:2023uhc}
Y.~Cai, M.~Zhu, and Y.-S. Piao, ``{Primordial black holes from null energy
  condition violation during inflation},''
  \href{http://arxiv.org/abs/2305.10933}{{\ttfamily arXiv:2305.10933 [gr-qc]}}.

\bibitem{Turner:1993vb}
M.~S. Turner, M.~J. White, and J.~E. Lidsey, ``{Tensor perturbations in
  inflationary models as a probe of cosmology},''
  \href{http://dx.doi.org/10.1103/PhysRevD.48.4613}{{\em Phys. Rev. D}
  {\bfseries 48} (1993) 4613--4622},
  \href{http://arxiv.org/abs/astro-ph/9306029}{{\ttfamily
  arXiv:astro-ph/9306029}}.

\bibitem{Hellings:1983fr}
R.~w. Hellings and G.~s. Downs, ``{UPPER LIMITS ON THE ISOTROPIC GRAVITATIONAL
  RADIATION BACKGROUND FROM PULSAR TIMING ANALYSIS},''
  \href{http://dx.doi.org/10.1086/183954}{{\em Astrophys. J. Lett.} {\bfseries
  265} (1983) L39--L42}.

\bibitem{LISA:2017pwj}
{\bfseries LISA} Collaboration, P.~Amaro-Seoane {\em et~al.}, ``{Laser
  Interferometer Space Antenna},''
  \href{http://arxiv.org/abs/1702.00786}{{\ttfamily arXiv:1702.00786
  [astro-ph.IM]}}.

\bibitem{Hu:2017mde}
W.-R. Hu and Y.-L. Wu, ``{The Taiji Program in Space for gravitational wave
  physics and the nature of gravity},''
  \href{http://dx.doi.org/10.1093/nsr/nwx116}{{\em Natl. Sci. Rev.} {\bfseries
  4} no.~5, (2017) 685--686}.

\bibitem{TianQin:2015yph}
{\bfseries TianQin} Collaboration, J.~Luo {\em et~al.}, ``{TianQin: a
  space-borne gravitational wave detector},''
  \href{http://dx.doi.org/10.1088/0264-9381/33/3/035010}{{\em Class. Quant.
  Grav.} {\bfseries 33} no.~3, (2016) 035010},
  \href{http://arxiv.org/abs/1512.02076}{{\ttfamily arXiv:1512.02076
  [astro-ph.IM]}}.

\bibitem{andrea_mitridate_2023_8106173}
A.~Mitridate and D.~Wright, ``Ptarcade,'' Apr., 2023.
\newblock \url{https://doi.org/10.5281/zenodo.8106173}.

\bibitem{Lamb:2023jls}
W.~G. Lamb, S.~R. Taylor, and R.~van Haasteren, ``{Rapid refitting techniques
  for Bayesian spectral characterization of the gravitational wave background
  using pulsar timing arrays},''
  \href{http://dx.doi.org/10.1103/PhysRevD.108.103019}{{\em Phys. Rev. D}
  {\bfseries 108} no.~10, (2023) 103019},
  \href{http://arxiv.org/abs/2303.15442}{{\ttfamily arXiv:2303.15442
  [astro-ph.HE]}}.

\bibitem{ellis_2020_4059815}
J.~A. Ellis, M.~Vallisneri, S.~R. Taylor, and P.~T. Baker, ``{ENTERPRISE:
  Enhanced Numerical Toolbox Enabling a Robust PulsaR Inference SuitE},'' Sep,
  2020.
\newblock \url{https://doi.org/10.5281/zenodo.4059815}.

\bibitem{enterprise-ext}
S.~R. Taylor, P.~T. Baker, J.~S. Hazboun, J.~Simon, and S.~J. Vigeland,
  ``{enterprise\_extensions},'' 2021.
\newblock \url{https://github.com/nanograv/enterprise_extensions}. v2.3.3.

\bibitem{Cai:2012va}
Y.-F. Cai, D.~A. Easson, and R.~Brandenberger, ``{Towards a Nonsingular
  Bouncing Cosmology},''
  \href{http://dx.doi.org/10.1088/1475-7516/2012/08/020}{{\em JCAP} {\bfseries
  08} (2012) 020}, \href{http://arxiv.org/abs/1206.2382}{{\ttfamily
  arXiv:1206.2382 [hep-th]}}.

\end{thebibliography}\endgroup
\bibliographystyle{utphys}

\end{document}